# Real GDP per capita in developed countries

Ivan O. Kitov

Abstract

Growth rate of real GDP per capita is represented as a sum of two components – a monotonically decreasing economic trend and fluctuations related to a specific age population change. The economic trend is modeled by an inverse function of real GDP per capita with a numerator potentially constant for the largest developed economies. Statistical analysis of 19 selected OECD countries for the period between 1950 and 2004 shows a very weak linear trend in the annual GDP per capita increment for the largest economies: the USA, Japan, France, Italy, and Spain. The UK, Australia, and Canada show a larger positive linear trend. The fluctuations around the trend values are characterized by a quasi-normal distribution with potentially Levy distribution for far tails. Developing countries demonstrate the increment values far below the mean increment for the most developed economies. This indicates an underperformance in spite of large relative growth rates.

Key words: economic development, economic trend, business cycle, GDP per capita

JEL classification: E32, O11, O57

## Introduction

Real economic growth has been studied numerically since Kuznets' works on accounting of national income and aggregate factor inputs. Hodrick and Prescott [1980] introduced a concept of two-component economic growth – an economic trend and a deviation or business cycle component. The trend component is responsible for the long-term growth and defines economic efficiency. In the long run, the deviation component of economic growth has to have a zero mean value. In 2004, Prescott and Kydland received a Nobel Prize for the study of "the driving forces behind business cycle" [Bank of Sweden, 2004], what demonstrates the importance of the best understanding of the growth processes and the explanation of the two-component behavior.

Prescott and Kydland, along with many other researchers, have proposed and studied exogenous shocks as the force driving fluctuations of real GDP growth rate. Their research during the last 25 years has revealed numerous features of principal variables involved in the description of the economic growth. There are many problems left in the theory of economic growth.

Kitov [2005a] proposed a model with GDP growth dependent only on the change in a specific age cohort in the population and the attained level of real GDP per capita. According to this model, real GDP per capita has a constant growth increment and the observed fluctuations can be explained by the population component change. In developed countries, real GDP per capita has to grow with time along a straight line, if no large change in the specific age population is observed. Relative growth rate of real GDP per capita has to be an inverse function of the attained level of real GDP per capita with a potentially constant numerator for developed economies. The paper is devoted to validation of the model using GDP per capita and population data for some selected developed countries. Our principal purpose is to demonstrate the possibility to decompose GDP per capita growth into the two components.

## 1. The model and data

Kitov [2005a] has developed a model explaining the observed real GDP growth rate variations in the USA. He has distinguished two principal sources of the per capita GDP growth in the USA – the change in 9-year old population and the economic trend related to the measured GDP per capita level. The trend has the simplest form – no change in absolute growth (annual increment) values and is expressed by the following relationship:

$$dG/dt=A \quad (1)$$

where $G$ is the absolute value of real GDP per capita, $A$ is a constant. The solution of this equation is as follows:

$$G(t)=At+B \quad (2)$$

where $B=G(t_0)$, $t_0$ is the starting time of the studied period. Hence, evolution of real GDP per capita is represented by a straight line if the second factor of growth has no cumulative effect. As discussed below, only some developed countries are characterized by a significant influence of the second factor.

Then, relative growth rate can be expressed by the following relationship:

$$dG/Gdt=A/G(t) \quad (3)$$

Relationship (3) indicates that the relative growth rate of per capita GDP is inversely proportional to the attained level of real GDP per capita, i.e. the observed growth rate should asymptotically decay to zero with increasing GDP per capita. On the other hand, the lower the level, the higher the growth rate. This inference might be a potential explanation for the concept of economic convergence. Relative growth rate must be higher in less developed countries, but the observed absolute gap in GDP per capita can not be overcome in future [Kitov, 2005b] unless some non-economic forces will disturb current status quo.

When considering real GDP per capita, one has to bear in mind the importance of a correction to be applied to the per capita GDP values related to the difference between the total population and population of 15 years of age and above, as discussed by Kitov [2005a]. Only this economically active population should be considered when per capita values are calculated. By definition, Gross Domestic Income, which is equivalent to GDP, consists of the personal incomes obtained by the population of 15 years of age and over and corporate income, the corporations owned by the same population category. Thus, one can treat the published (original) readings of GDP per capita as biased and to be corrected for (multiplied by) the corresponding population ratio, i.e. the ratio of the total population and the population above 14 years of age.

Figure 1 shows the population ratio as obtained from the OECD population data [OECD, 2006]. When absent the missed readings are substitute with those for the closest year from above. Between 1955 and 2003, the ratio is characterized by an overall decrease with a slight increase demonstrated by some countries in the 1960s and 1970s. Currently, all the countries have the ratio below 1.3. In the 1950s, the ratio was above 1.3 for all the countries except Austria and Belgium. The last country met the decrease is Ireland - the drop started in 1980. Italy has had the lowermost ratio since 1970.

The decreasing ratio implies that the GDP per capita readings during the period between 1950 and 1970 are underestimated compared to those during the last 35 years. The larger is the total drop in the ratio during the entire period of the observation, the larger is the overall correction. In the study, the original and the corrected per capita GDP values are used and compared.

A cross-country comparison implies that GDP per capita is measured in the same currency units. There are two principal possibilities to reduce national readings of GDP per capita to some common scale: to use currency exchange rates or purchase power parities. In the study, we use the latter approach and data provided by the Conference Boars and Groningen Growth and Development Center [CB GGDC, 2006]. For developed countries, two estimates of GDP per capita level are available: measured in 2002 US dollars, for which "EKS" purchasing power parities have been used [CB GGDC, 2005], and that expressed in 1990 US dollars, with the conversion at “Geary-Khamis” PPPs. These PPPs are obtained from the Organization for Economic Co-operation and Development [OECD, 2005]. Being an improvement on the previous dataset, the “EKS” PPPs are considered as more accurate and reliable. Amplitude of the change induced by the transition from “Geary-Khamis” PPPs to those of “EKS” is evaluated for the counties under investigation. This change potentially characterizes uncertainty in the GDP per capita readings obtained with the PPP approach.

Only nineteen from thirty OECD member countries are analyzed. The selected countries meet some general criteria: 1) large economy size denominated in dollars; 2) continuous observations during the period between 1950 and 2003; 3) high level of real GDP per capita. According to the size criterion, small economies like Iceland and Luxembourg were excluded. When applied, the second criterion rejects Germany from the consideration. The third criterion has excluded such countries as Turkey, Poland and other new EU members. Finland and Korea have been excluded from the analysis with no reason at all.

Figure 2 illustrates the variations induced by corrections made for the “EKS” PPPs compared to those of “Geary-Khamis”. The original values of the mean increment of GDP per capita (“EKS” and “Geary-Khamis”) for every country are normalized to the corresponding values for the USA. The normalized values are consistently higher for the "EKS" PPPs, i.e. the GDP per capita values converted at "Geary-Khamis" PPPs were underestimated for all the countries. The difference varies with country and reaches 5% to 7% for Austria, Norway and Ireland. For the largest developed economies, the mean increments of GDP per capita expressed in 2002 dollars converge to that for the USA. We use the GDP per capita readings expressed in 2002 US dollars are used in the study. The only exception is the statistical description of the observed fluctuations.

Figure 3 displays the averaged values of the annual GDP per capita increments denominated in 2002 US dollars for the period between 1950 and 2003. The original and corrected for the population ratio values for the nineteen countries are normalized to the corresponding values for the USA. As before, this procedure allows a homogeneous comparison of the mean values. The corrected normalized values can be lower or higher than those for the original set. The sign of the change depends on the overall behavior of the population ratio during the entire period compared to that for the USA. Ireland, Austria and Norway are excellent examples of the originally underestimated GDP per capita values. Canada, Italy and Spain demonstrate an opposite behavior.

It is worth noting that the correction for the population ratio is of lower magnitude than that induced by the transition from the “Geary-Khamis” PPP to “EKS” one. The population correction is important, however, because it reduces potential uncertainty in the decomposition of the GDP per capita growth into the two components. The purchase power parity approach to the estimation of national GDP also needs some further improvements. Magnitude of the difference between the GDP per capita values converted at "Geary-Khamis" and "EKS" PPP sets is too high to believe that all the problems with the homogeneous and accurate cross-country comparison are resolved.

## 2. The GDP per capita trend

The nineteenth selected countries are presented in alphabetic order. Figure 4 shows the evolution of the annual increment of real GDP per capita for Australia as a function the GDP per capita level itself for the both original and corrected GDP per capita values. This is a natural representation for relationship (1). The population corrected values are connected by a solid line in order to illustrate the evolution in time. Open circles represent the original readings. In addition, three straight lines are drawn in the Figure. Bold line corresponds to the mean increment value of the population corrected GDP per capita for the entire period between 1950 and 2003. Being a constant, this line is parallel to x-axis. The second and the third (solid) lines represent two linear regressions corresponding to the original and corrected data sets. Relationships for the regressions are also shown in the Figure, the lower one always associated with the original GDP per capita set.

The model presented in section (1) implies that the mean value line has to coincide with the linear regression line, if the population induced component has a zero mean value. The observed fluctuations of the GDP per capita annual increment can be either predetermined or random depending on the underplaying population change characteristics. In terms of statistics, one can expect a normal distribution of the population change. The number of processes affecting birth rate, mortality rate and international migration processes is very large and according to the central limit theorem this leads to an approximately normal distribution. However, random fluctuations of population do not mean an unpredictable economic growth rate. For example, the number of nine-year-olds in the USA can be counted with any desirable accuracy and completely define observed economic growth. Statistical features of the GDP per capita increment will be discussed later in this section.

Australia demonstrates a divergence between the regression lines and the mean value line. A positive linear trend has to indicate a more intensive growth of the specific age population in recent years compared to that in the 1950s and 1960s. This effect is observed also for the other

English-speaking countries under investigation. The average increment is $452. The largest deviation from the mean is -$1113. The linear trend coefficient is lower for the corrected data set than that for the original set. This is a common feature for almost all the studied countries. Table 1 lists the mean values and regression coefficients for all the data sets: the original and population corrected, converted at "EKS" and "Geary-Khamis" PPPs.

Alphabetically, the following country is Austria. As for Australia, Figure 5 displays the GDP per capita increment values and the corresponding mean value and linear regression lines. The average increment value for Austria is $548. This value is well above that for Australia. A prominent feature is an almost horizontal regression line for the population corrected data set with the linear trend coefficient $6*10^{-5}$. Effectively, the mean line and the regression lines coincide, as predicted by relationship (1). For the original set, the coefficient is 0.0041 and some positive trend is observed. From the Figure, one can conclude that the specific age population has not changed during the last 53 years and its fluctuations were self-compensating at the short-run during these years. The largest fluctuation amplitude relative to the mean value is $792. In relative terms, such a deviation from the mean value is almost 4%.

Figure 6 presents Belgium. The average increment is $483 and the regression lines are characterized by a positive linear trend that is higher for the original readings. By these characteristics, Belgium is closer to Australia than to Austria. The largest deviation from the mean value is -$956 or -5%. The negative deviations are sharp and deep compensating longer periods of weaker growth. The last twenty years have been relatively successful for Belgium. One can expect a compensating decrease, as was observed between $22000 and $25000.

Results for Canada are shown in Figure 7. This country is similar to Australia, but is characterized by a lower mean increment ($425) and lower trend coefficients. An important feature of the Canadian economy fluctuations is their amplitude reaching $1650 – the highest for the studied economies. The two deep drops at $26000 and $31000 compensate a relatively successful history during the rest of the period.

Figures 8 and 9 present per capita GDP for Denmark and France. These countries are similar in terms of a weak linear trend, positive for Denmark and negative for France, and close values of the mean increment - $465 and $483 respectively. If to neglect the trend, one can conclude that the observed fluctuations are random and characterized be a zero mean value. Because of a limited time period of the observation, the trend values can be also affected by side or truncation effects. The fluctuations are of different character for various countries and different periods of observations are necessary for suppressing the side effects. For a majority, side effects are weak and results in very small linear regression coefficients.

Figure 10 represents Greece. The Greek economy had some hard years in its history and is characterized by a relatively low mean increment value - $390. There were some bright periods, however, including the last five to seven years, but the overall performance, expressed also in the low linear trend value, does not indicate a cloudless future. This country can not be used as an example of a developed economy maximizing its performance over years.

An opposite example of an excellent recovery gives Ireland with corresponding results displayed in Figure 11. A slow start was quickly compensated and the last twenty years of an extremely fast growth resulted in the leading position in the world economy with the mean increment $678. There are some doubts, however, that future will be so successful. Such a long and quick growth always ends up in a depression. This was observed in Japan and is related to the long-term decrease in the number of the specific age population [Kitov, 2005a]. Ireland has managed to increase birth rate for a very long period and has an age structure similar to that observed in Japan 20 years ago. The population distribution is currently peaked near 20 years with the defining age of 18 years. The years to come will demonstrate only decrease in the defining age population.

The next three countries are Italy, Japan, and Netherlands. Results for them are represented in Figures 12 through 14 and are similar to those for Belgium and France – a weak positive or negative linear trend and the mean increment near $450. Japan has the mean value of $538, however. These are also good examples of a zero linear trend in the history of GDP per capita increment.

Norway and New Zealand are represented in Figures 15 and 16, respectively, and are very similar the pair Ireland/Greece. From the point of view of the current study they do not provide any additional insight into the GDP increment behavior.

Portugal (Figure 17) is between Greece and New Zealand. Spain (Figure 18) and Sweden (Figure 19) are similar to other large European economies with a weak linear trend of the per capita GDP increment and the mean value around $450. Switzerland (Figure 20) had a decreasing increment which can be potentially explained by a permanent decrease in the young population portion.

The UK and USA differ only in the mean increment value: $444 and $557, respectively. Positive linear trend is relatively high for the both countries. The US trend is well explained by the change in the nine-year-old population [Kitov, 2005a]. When corrected for the integral nine-year-olds change between 1950 and 2003, the US mean value is only $462, i.e. in the tight group of the largest economies. The UK statistical agencies do not provide accurate population estimates for the entire period, but from the mean value one can assume that there was no significant increase in the number of nine-year-olds.

## 3. Discussion and conclusions

The nineteen countries show various behavior of GDP per capita during the period between 1950 and 2003. There are countries with a slightly negative trend of GDP per capita increment: France, Italy, Switzerland, and Japan. Despite the common negative trend the countries have quite different mean increments: $483, $458, $538, and $398 respectively. Austria is characterized by a zero trend and has a large mean value $548. One can count it into the club.

A majority of European countries including Belgium, Denmark, Greece, Netherlands, Portugal, Spain, and Sweden are characterized by a slightly positive trend. The mean increment value varies, however, from of $347 for Portugal to $483 for Belgium. Greece and Portugal showed a weak growth in the beginning of the period, but have recovered to a normal pace. There are two outstanding European countries – Ireland and Norway. Their mean increment is very high, but the countries have a strong downward tendency during the last three to five years. One can expect them to follow the path of Japan – from a strong growth to a long period of stagnation. At the same time, the countries are small. Their influence on the world economy is negligible. Thus, we also deny the countries to influence our analysis of economic trend.

The studied English speaking countries are characterized by a large positive trend, but should be separated into two groups. The first consists of only one member – New Zealand. The principal characteristic is a very poor performance during the period. The second includes Australia, Canada, the UK, and the USA. The mean increment for them is between $424 and $462. For the USA, the value is obtained with the correction of the specific age population total growth. The accurate population estimates available for the USA allowed explanation for not only the trend, but also the largest fluctuations. Smaller deviations from the mean value are compatible with the characteristic noise of the population estimates and are not so well correlated.

The mean increment varies with country and at most reaches $548 in Austria. There is no specific reason for Austria to grow so fast except the same as for Ireland – neighboring a very powerful economy. The Austrian influence on the world economy is also very limited.

The mean increment in the USA corrected for the total nine-year-olds change between 1950 and 2003 equal to 0.82 is only $557*0.82=$462 [Kitov, 2005a]. For France, this factor is 0.97. The author failed to find reliable data for the other countries under study. The most popular mean increment lays between $425 for Canada and $483 for Belgium (the French value is $483*0.97=$469). We do not consider the countries with known political and economic problems in past – Greece, Portugal, New Zealand. Overall, they demonstrated consistent underperformance. Switzerland surprisingly joins the club of the countries with weak growth, but the reason might be of a different nature – the decreasing population of the defining age.

The above analysis has revealed that the largest developed economies are characterized by very close values of the mean GDP per capita increment for the period between 1950 and 2003. The mean income defines the long-term economic trend. Thus, the countries are characterized by the same trend level not depending on the attained level of GDP per capita. A different question, what are statistical properties of the residual growth – fluctuations? In order to answer the question frequency distributions in $200 wide bins were constructed for each of the original and population corrected data set.

Figure 23 shows the frequency distribution for the original GDP per capita readings as obtained using “EKS” PPPs. Amplitude of the fluctuations is measured from the corresponding mean value for each of the nineteen countries. The distribution is very close to a normal one with the mean value and standard deviation obtained from the original data set - $0 (the mean value is subtracted) and $359, respectively. Corresponding normal distribution is shown by open circles. The tails of the real distribution are above the predicted values of the normal distribution. This effect is often observed in natural sciences and is associated with inaccurate measurements, limited amount of readings, and sometimes with action of some real factors. One can also suggest a Levy distribution as better presenting the observed tails.

Figure 24 displays the same curves for the population corrected GDP per capita values. Due to relatively narrower bins ($200 original not equal to $200 corrected for the population ratio) the actual distribution is characterized by higher deviations from the normal distribution. At the same time, the central part of the actual distribution is still very close to the normal one. One can conclude that the correction for the population ratio does not improve the fluctuation distribution in terms of theoretical representation.

Figure 25 represents the case of “Geary-Khamis” PPPs. The GDP per capita increment values are smaller and the actual frequency distribution is very close to the corresponding normal distribution, especially in the central zone. One can assume that these PPPs are more consistent with the normal distribution of the fluctuations than those of the “EKS”. There is no ultimate requirement for the fluctuations to be normally distributed.

The nineteen countries can be also separated into two large groups according to the trend and mean increment value considering potential bias introduced by changing population and economic failures. In Figure 26, for the reasons considered above in the text, we excluded the following economies: Greece, Ireland, Norway, New Zealand, Portugal, and the USA. There is a slight improvement on the results for the full data set.

Hence, a successful large-scale economy might be characterized by two principal features: real per capita GDP increment randomly fluctuates around some constant, and the constant is around $450 (2002 US dollars). It is very probable that the fluctuations are normally distributed. This hypothesis is strongly supported by the above observations. There are numerous possibilities to improve convergence of the results if to obtain accurate population data and to enhance the PPP conversion procedure. The mean increment value $450 is a good starting point for calibrating the PPP methodology and evaluation of long term economic performance for developed countries.

Developing counties also can be evaluated according to their compliance to the principal characteristic for developed countries. One may often hear about a “fast” growth of some developing countries like China and India. There is not criterion, however, to compare their growth rate to that expected in the USA, for example, at the same level of economic development. Using the mean increment, one can easily estimate a pace for any developing country compared to that observed in the developed world. For China, India, and the USSR, the increment evolution compared to that for France is represented in Figure 27. One can see that the countries demonstrate the per capita GDP increment far below the French mean value (1990 dollars are used as only available for all the countries). Having an intention to catch up a developed economy, any developing country has to analyze its time history of the GDP per capita increment [Kitov, 2005b]. No deficiency has to be allowed on the way to prosperity because any gap is created forever judging from the history of such successful developed countries as the USA, France, and others.

## References

The Conference Board and Groningen Growth and Development Centre (2005), Total Economy Database, August 2005, http://www.ggdc.net

Hodrick, R and E. Prescott (1980), Postwar U.S. business cycles: an empirical investigation, Discussion Paper, Northwestern University

Kitov, Ivan, (2005a) “GDP growth rate and population”, Research Announcements, Economics Bulletin, Vol. 28 no. 9 p. 1.

Kitov, Ivan, (2005b) "Modelling the transition from a socialist to capitalist economic system," Research Announcements, Economics Bulletin, Vol. 28 no. 11 p. 1.

The Bank of Sweden Prize in Economic Sciences in Memory of Alfred Nobel 2004, Press Release, 11 October 2004, http://nobelprize.org/economics/laureates/2004/press.html

Organization for Economic Co-operation and Development (2005), Purchasing Power Parities 2002, Paris, January 2005, http://www.oecd.org/department/0,2688,en_2649_34357_1_1_1_1_1,00.html

Organization for Economic Co-operation and Development (2006), Corporate Data Environment, Labor Market Statistics, DATA, User Queries, January 8, 2006, http://www1.oecd.org/scripts/cde/

## Tables

Table 1. Mean values of GDP per capita increment for the original and population corrected readings in 1990 $ (converted at Geary-Khamis PPPs) and 2002 $ (“EKS” PPPs). Coefficients of linear regression (trend) for the original and corrected GDP per capita values denominated in 2002 $.  Negative trend values are highlighted.

| | original, 2002 $ | corrected, 2002 $ | trend original | trend corrected | original, 1990 $ | corrected, 1990 $ |
|---|---|---|---|---|---|---|
| Australia | 386 | 452 | 0.0224 | 0.0261 | 300 | 351 |
| Austria | 464 | 548 | -6 E(-5) | 0.0041 | 328 | 387 |
| Belgium | 407 | 483 | 0.0061 | 0.0090 | 297 | 353 |
| Canada | 384 | 425 | 0.0086 | 0.0135 | 302 | 335 |
| Denmark | 396 | 465 | 0.0098 | 0.0080 | 300 | 353 |
| France | 404 | 483 | -0.0041 | 0.0005 | 304 | 364 |
| Greece | 332 | 390 | 0.006 | 0.0113 | 220 | 258 |
| Ireland | 548 | 678 | 0.0608 | 0.0665 | 400 | 495 |
| Italy | 407 | 459 | -0.0049 | -0.0015 | 295 | 332 |
| Japan | 476 | 538 | -0.0066 | -0.0070 | 367 | 415 |
| Netherlands | 402 | 462 | 0.0030 | 0.0054 | 291 | 335 |
| Norway | 545 | 666 | 0.0325 | 0.0307 | 384 | 470 |
| New Zealand | 229 | 256 | 0.0176 | 0.0173 | 172 | 192 |
| Portugal | 302 | 348 | 0.0074 | 0.0131 | 223 | 257 |
| Spain | 394 | 449 | 0.0097 | 0.0188 | 273 | 311 |
| Sweden | 370 | 436 | 0.0087 | 0.0091 | 280 | 330 |
| Switzerland | 358 | 398 | -0.0256 | -0.0187 | 247 | 275 |
| UK | 370 | 444 | 0.0223 | 0.0231 | 270 | 324 |
| USA | 470 | 557 | 0.0174 | 0.0177 | 371 | 439 |

## Figures

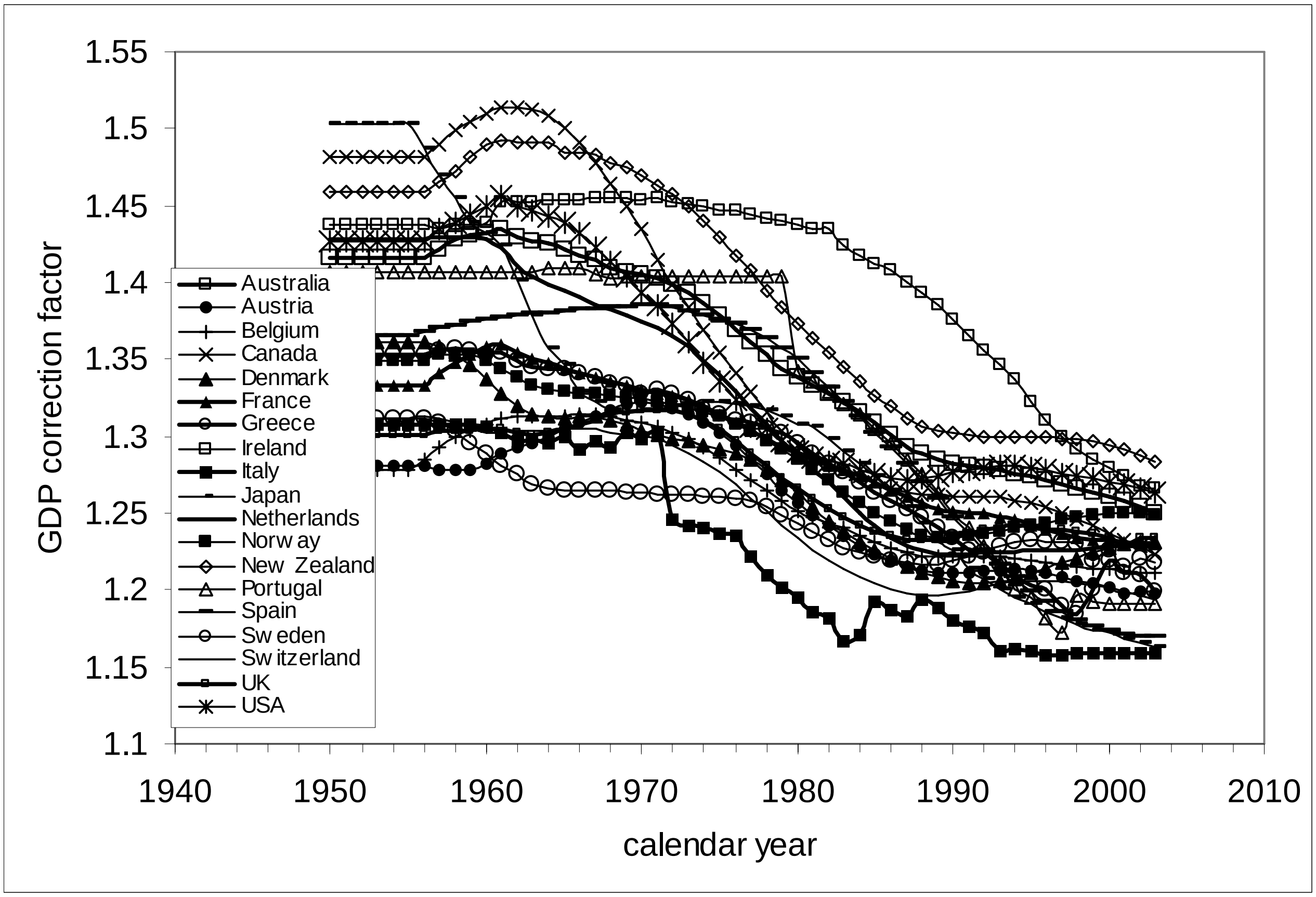

Fig. 1. Evolution of a ratio of the total population and the population above 15 years of age for the selected OECD countries [OECD, 2006]. High values of the ratio mean a relatively underestimated real GDP per capita and vice versa. A general feature of the curves is that after a small increase observed for some countries in the 1960s and 1970s the ratio decreases into the range between 1.3 and 1.15 in 2000. Thus the earlier GDP per capita values are relatively underestimated and the later readings are relatively overestimated. The longest period of a high ratio is observed in Ireland. Italy has a consistently low ratio. In the USA, the ratio drops from 1.45 in 1960 to 1.27 in 2003.

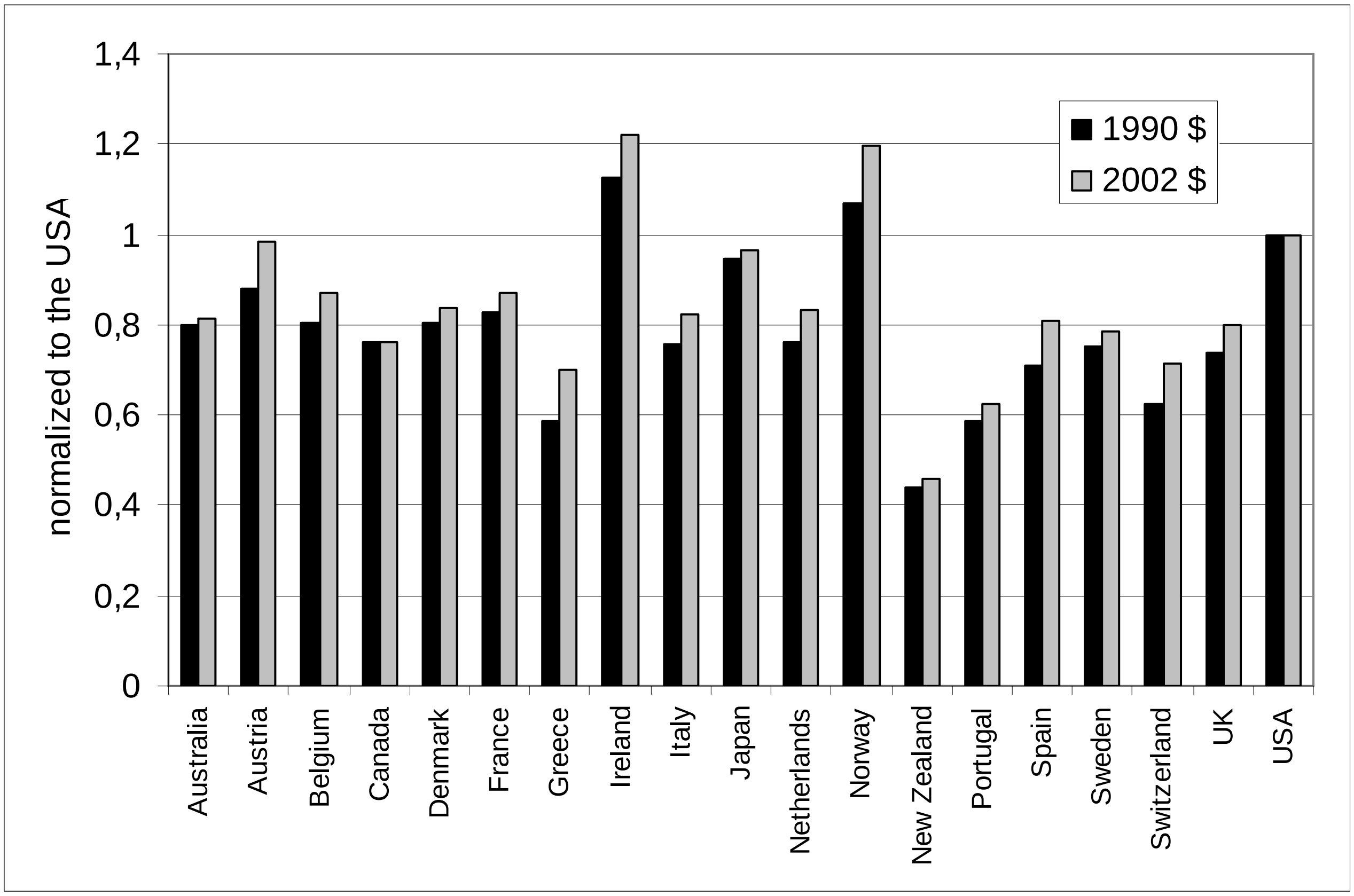

Fig. 2. Comparison of two real GDP per capita data sets denominated in 1990 and 2002 dollars [CB and GGDC, 2005], for which "Geary-Khamis" and "EKS" purchasing power parities have been used respectively. Original values of the mean GDP per capita increment are normalized to the corresponding value for the USA. The normalized values are consistently higher for the "EKS" PPPs (except Canada), i.e. the GDP values converted at "Geary-Khamis" PPPs were underestimated for all the countries. The difference varies with country, however, and is larger than that between the original and corrected for the population values presented in Figure 3.

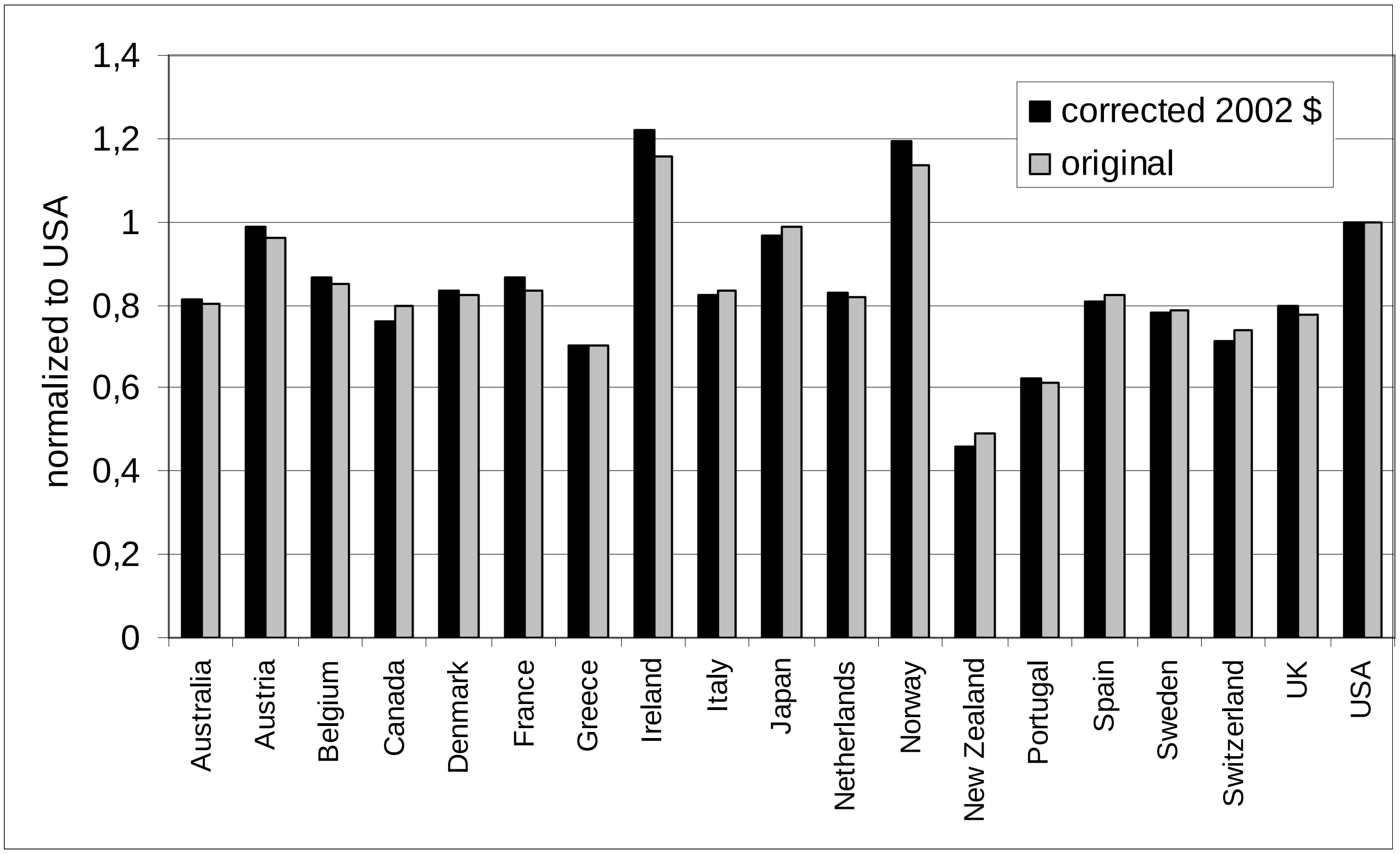

Fig. 3. Comparison of the original and corrected for population mean values of GDP per capita increment expressed in 2002 dollars converted at "EKS" PPPs [CB and GGDC, 2005]. The values are normalized to the corresponding value for the USA for a homogeneous representation. There are countries with overestimated (where the corrected value is below the corresponding original value) and underestimated (opposite) values relative to the USA.

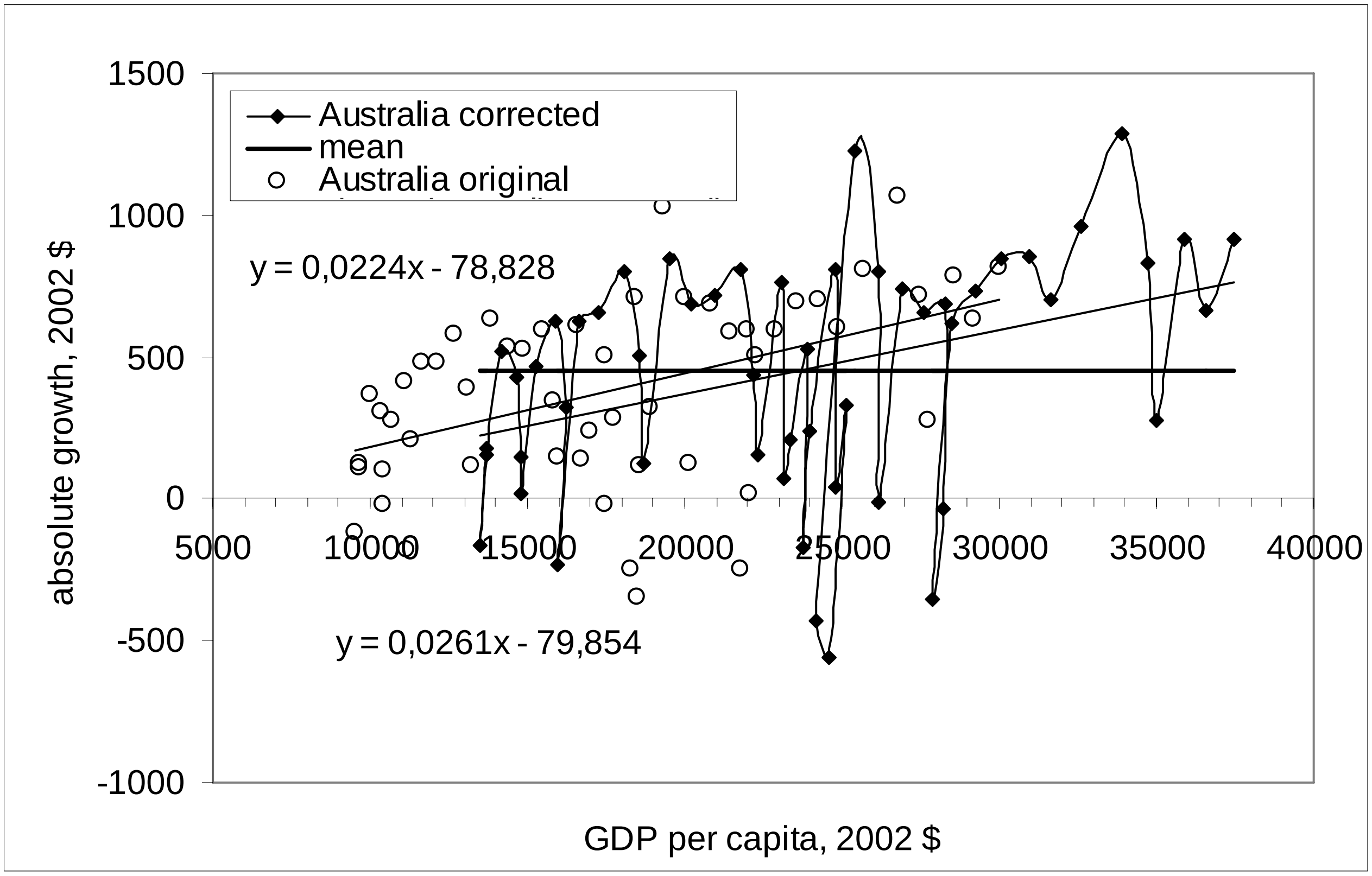

Fig. 4. Increment of real GDP per capita (2002 dollars) vs. real GDP per capita in Australia for the period between 1950 and 2004. Two sets are presented - the original (open circles) and corrected for population (filled diamonds). Consequent values of the latter set are connected by a solid line for illustration of the evolution in time. Bold line represents the mean value of $452 for the population corrected data set. Two solid lines show linear regressions lines. The corresponding linear relationships are displayed, the lower relationship being associated with the original data set. The regression straight lines differ from that for the mean value.

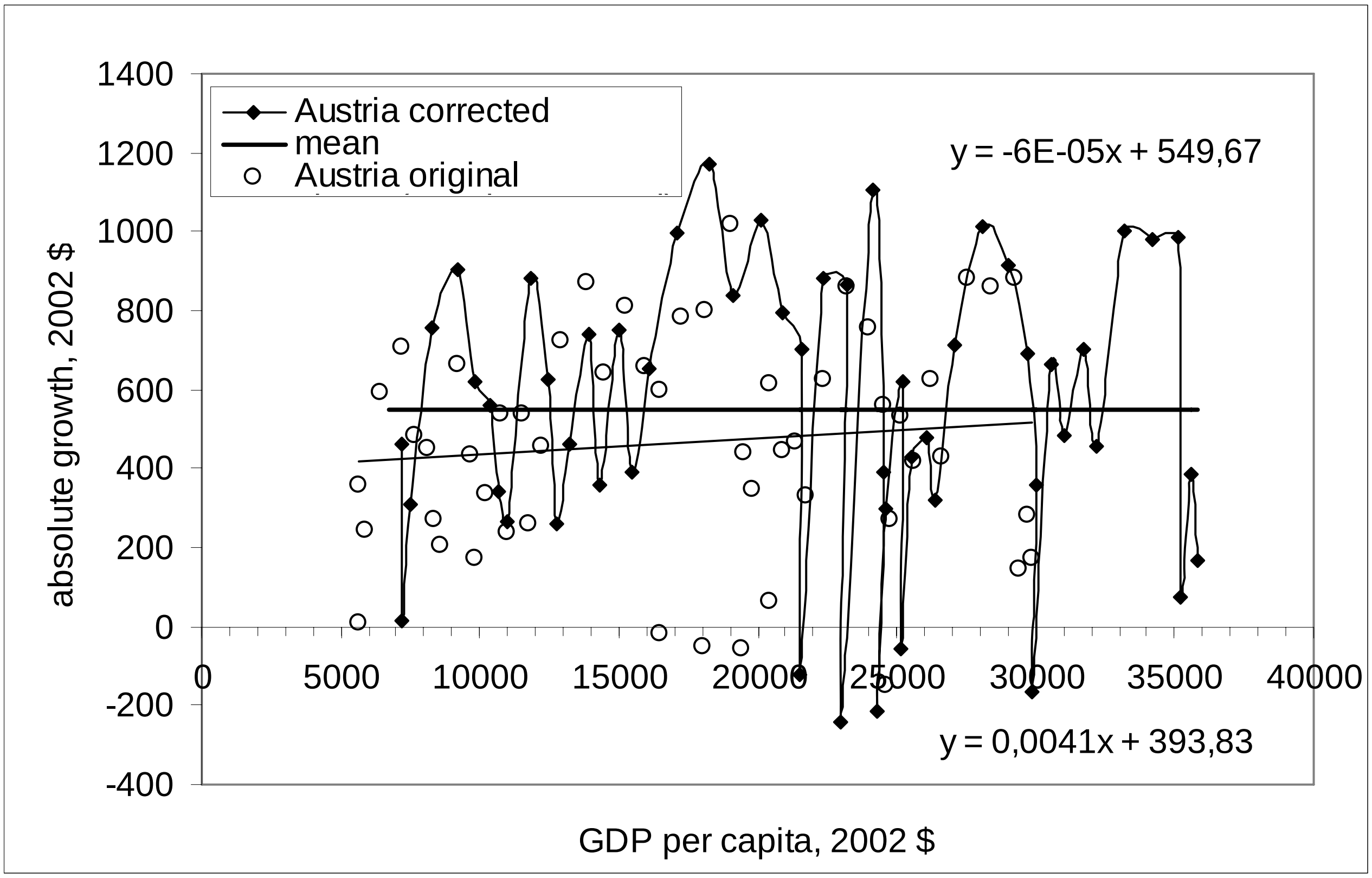

Fig. 5. Same as in Figure 4 for Austria. The mean value is $548. The linear regression line for the corrected GDP increment values almost coincides with that for the mean value. The original data set is characterized by a positive trend with a coefficient 0.004, i.e. $0.4 per $100.

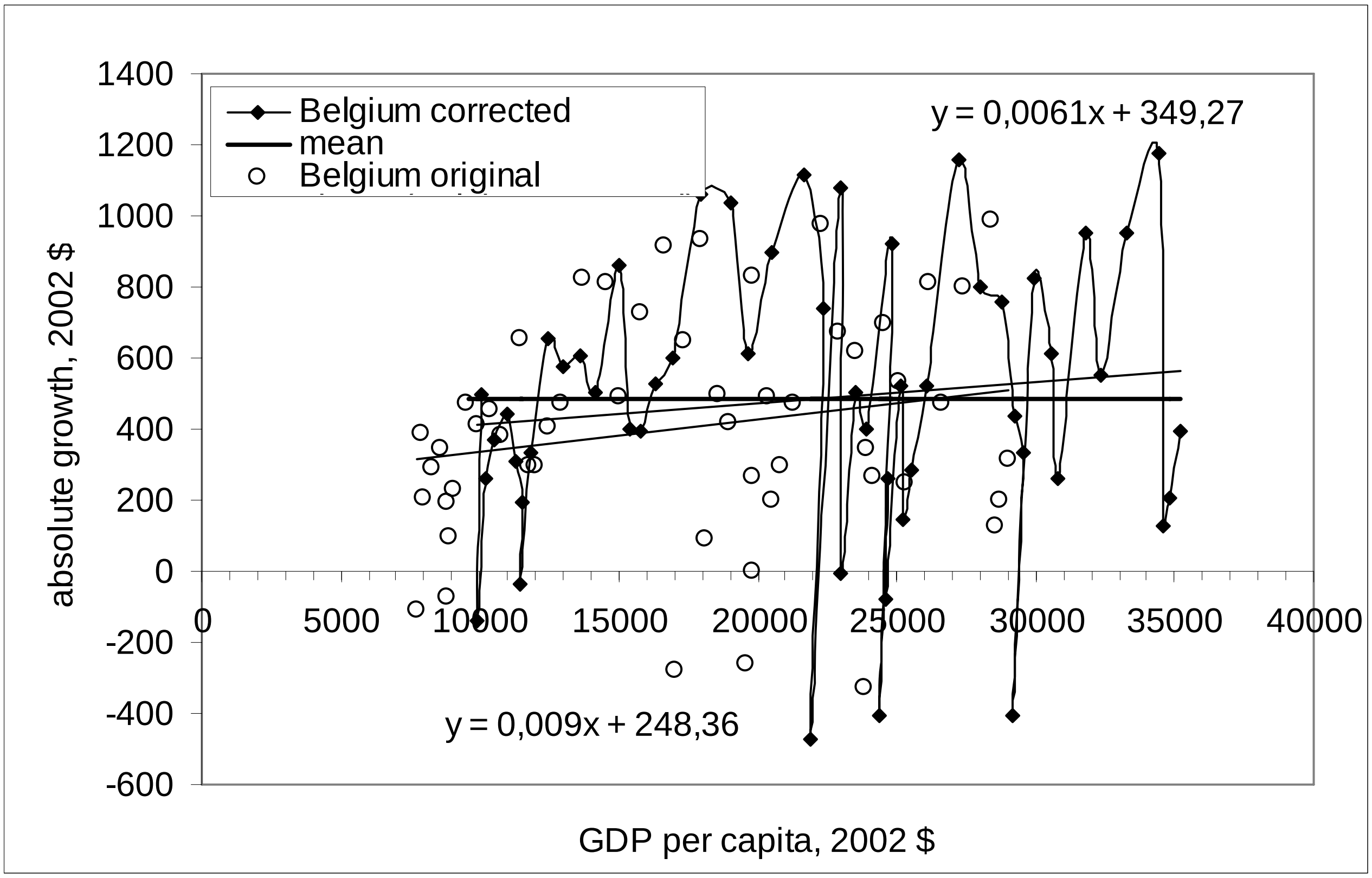

Fig. 6. Same as in Figure 4 for Belgium. The mean value is $483. The linear regression line for the corrected GDP values has a smaller positive coefficient than that for the original set, but higher than that for Austria. The mean value for Belgium is lower, however.

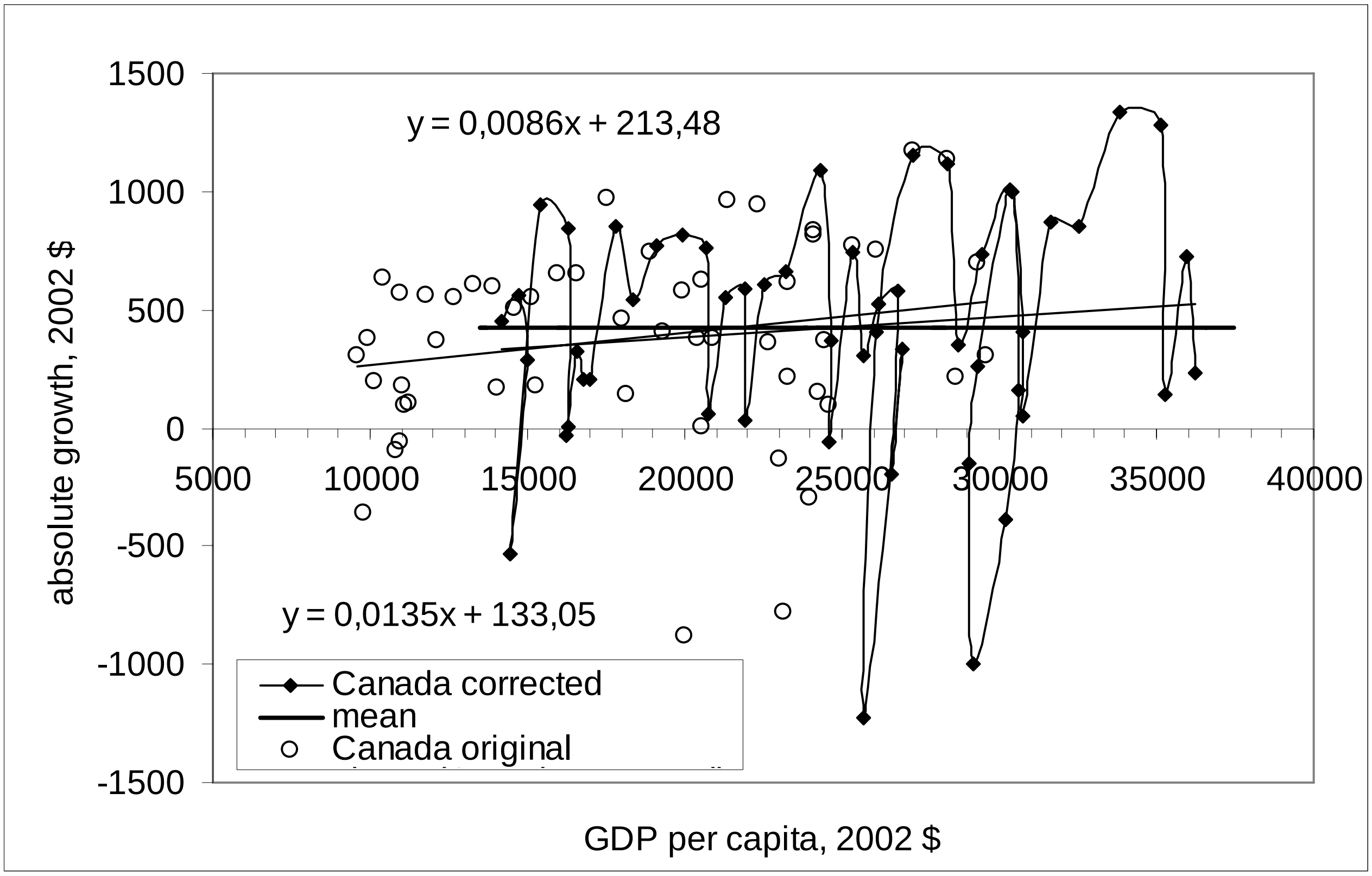

Fig. 7. Same as in Figure 4 for Canada. The mean value is $424. The linear regression line for the corrected GDP values has a smaller positive coefficient (0.0086) than that for the original set (0.0135).

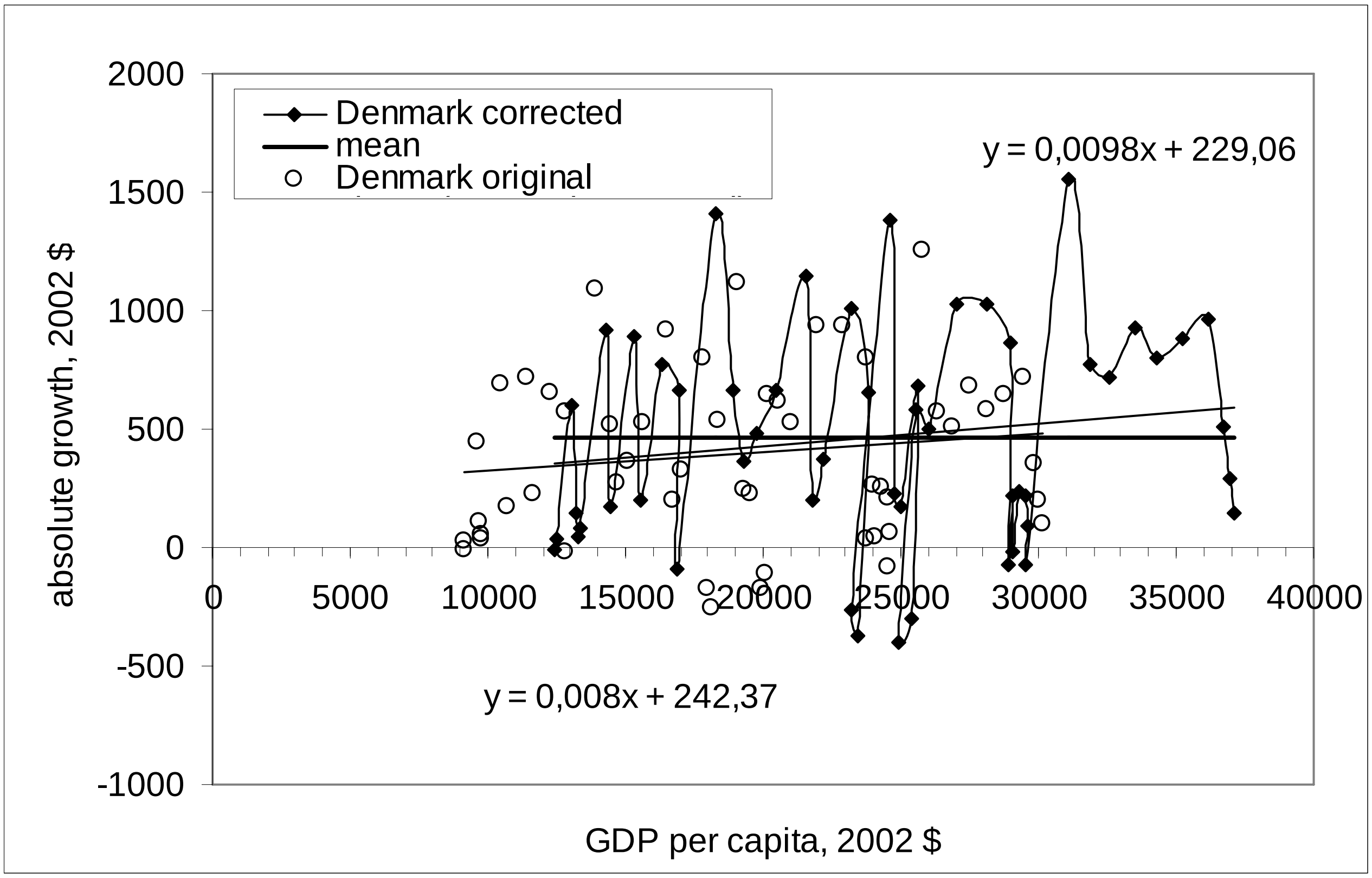

Fig. 8. Same as in Figure 4 for Denmark. The mean value is $465. The linear regression line for the corrected GDP values has a larger positive coefficient (0.0098) than that for the original set (0.008). This behavior differs from that for the four previous countries.

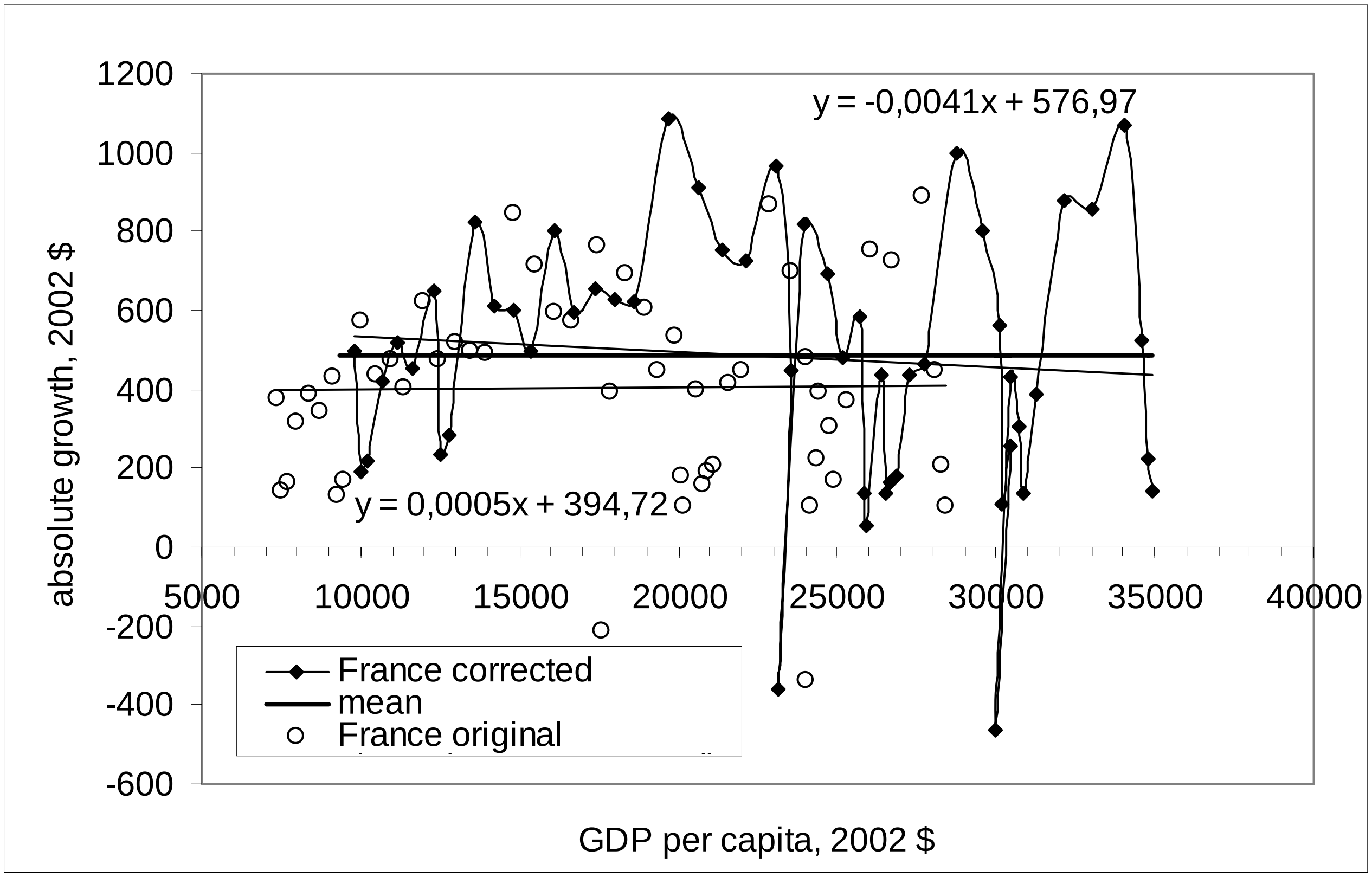

Fig. 9. Same as in Figure 4 for France. The mean value is $483. The linear regression line for the original GDP values is practically parallel to the mean value line. The line for the corrected data set is characterized by a negative trend coefficient (-0.004).

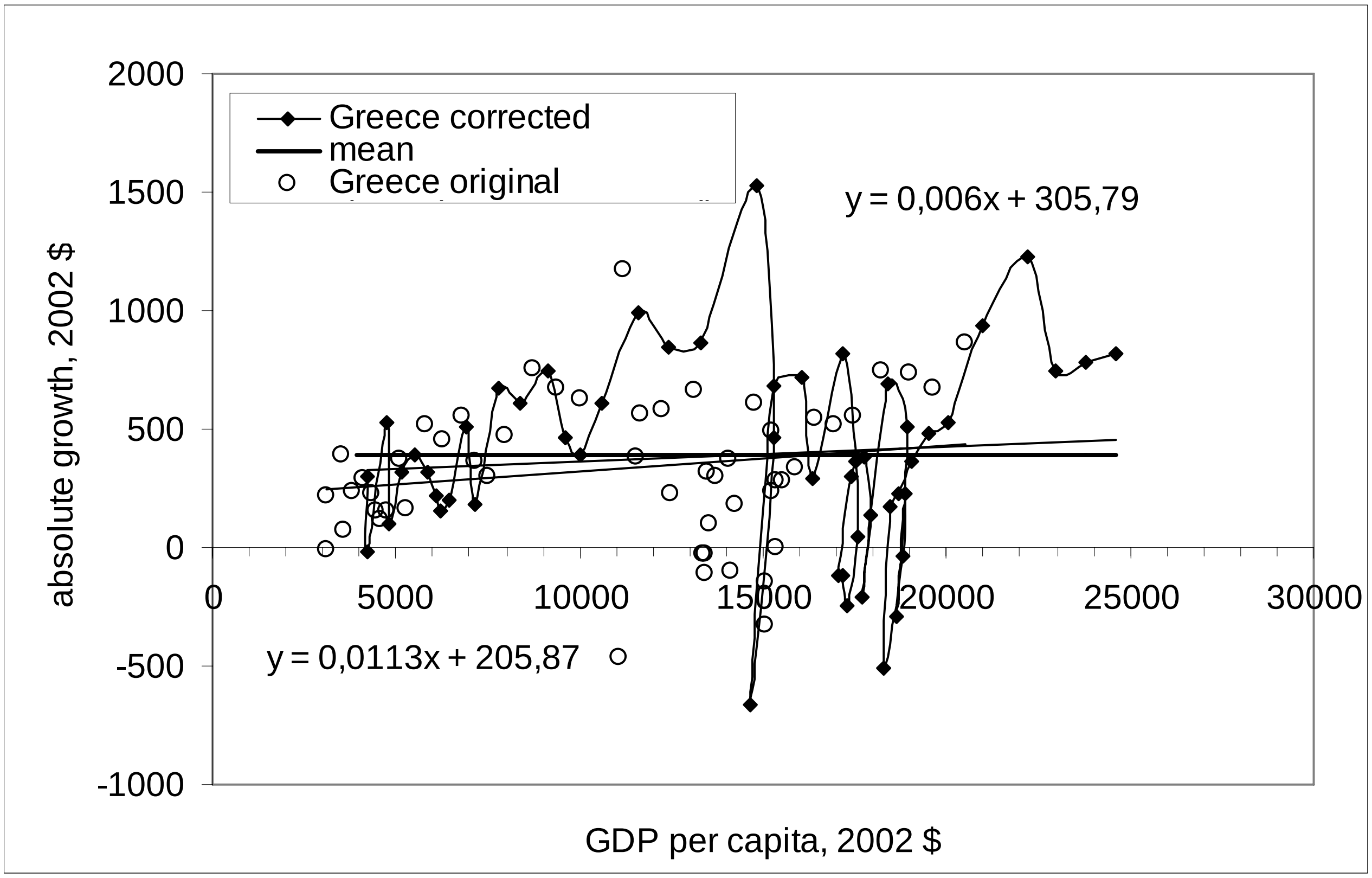

Fig. 10. Same as in Figure 4 for Greece. The mean value is as low as $390. The linear regression line for the corrected GDP values has a smaller positive coefficient (0.006) than that for the original set (0.0113).

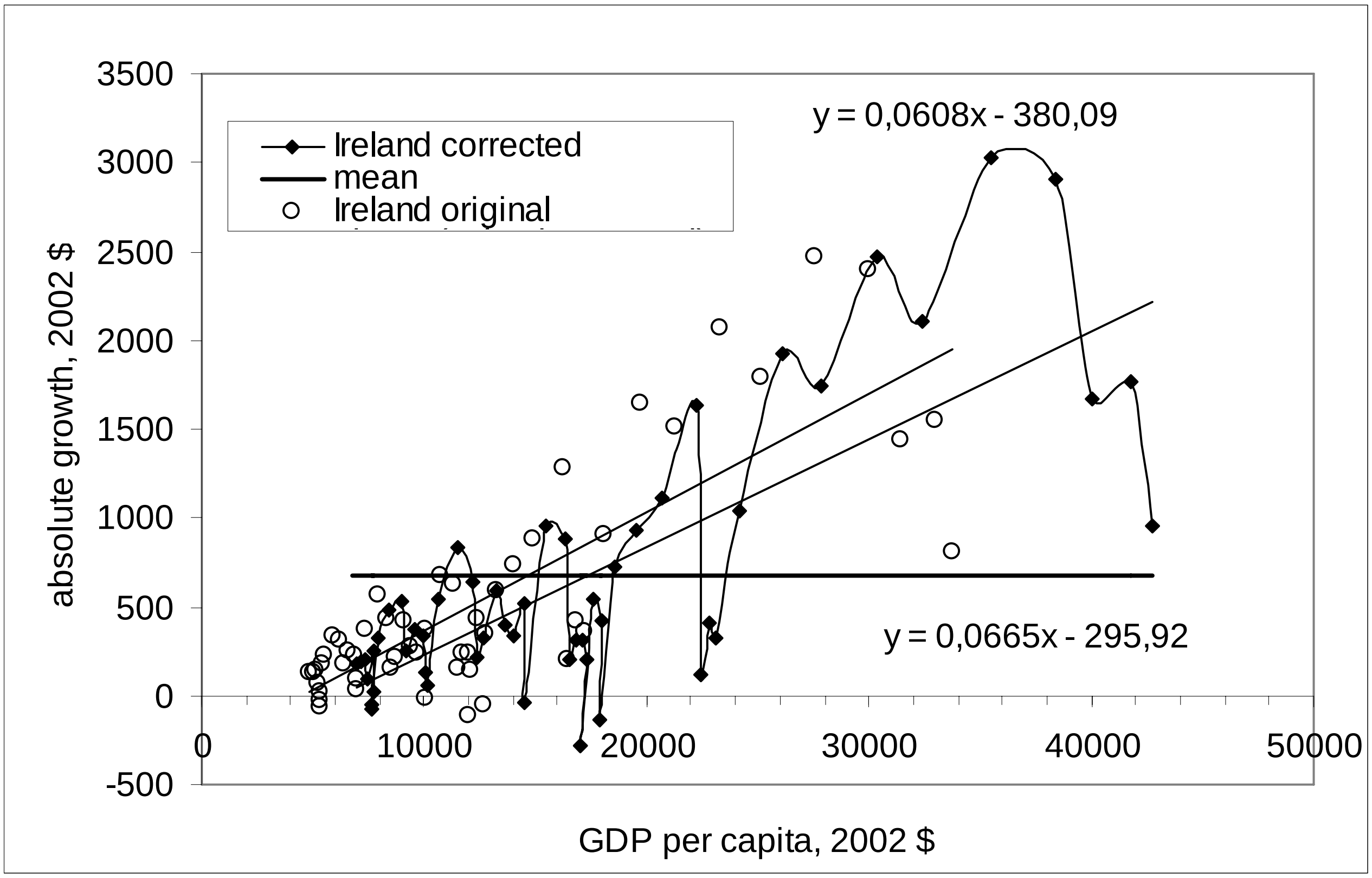

Fig. 11. Same as in Figure 4 for Ireland. The mean value is $678. The growth of the real GDP per capita is outstanding during the last twenty years. There is a downward tendency during the last four years, however.

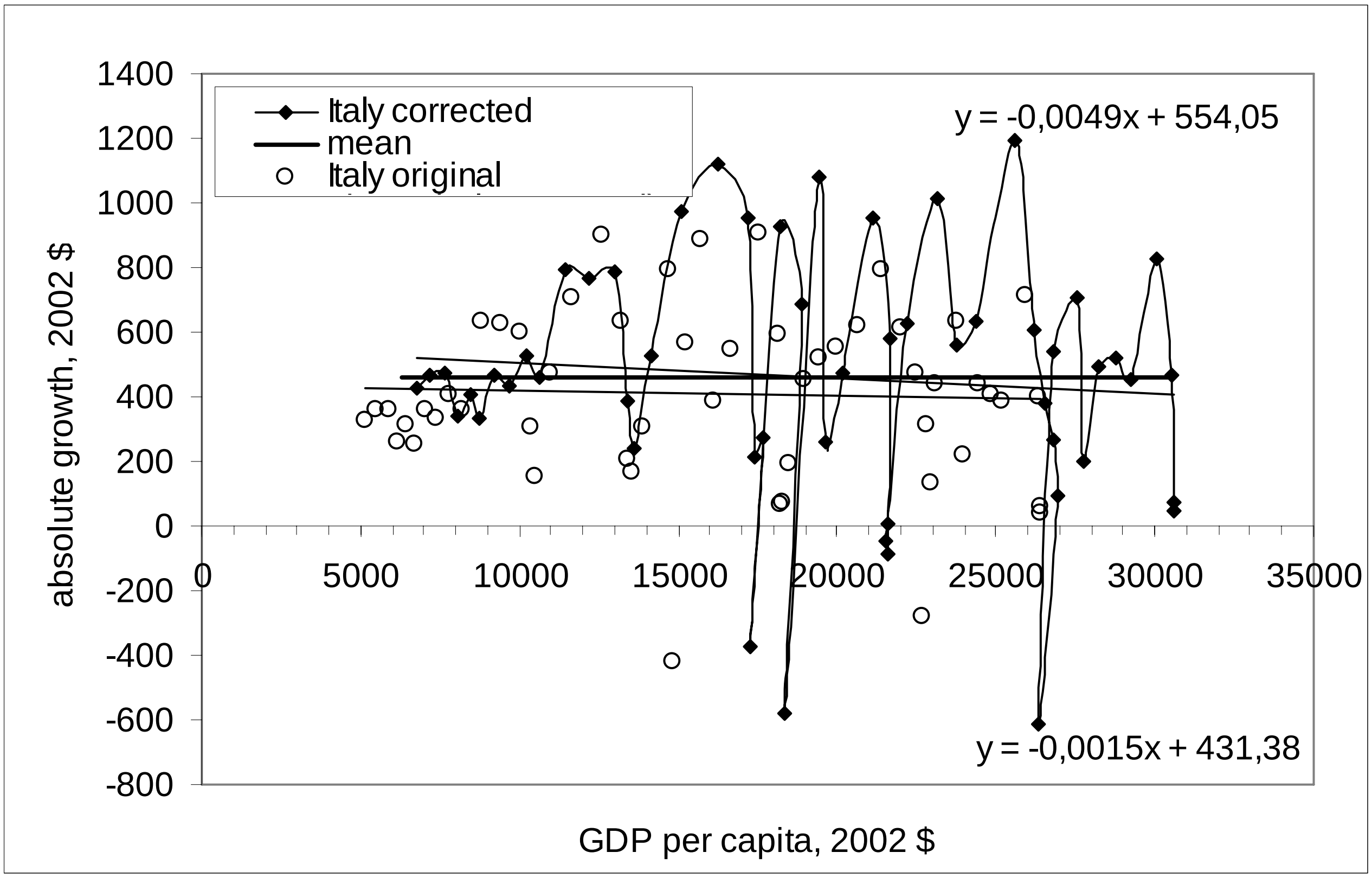

Fig. 12. Same as in Figure 4 for Italy. The mean value is $459. Both the corrected and original GDP values produce a negative trend, the former being of a larger absolute value. Nevertheless, the lines are very close to those for the mean values.

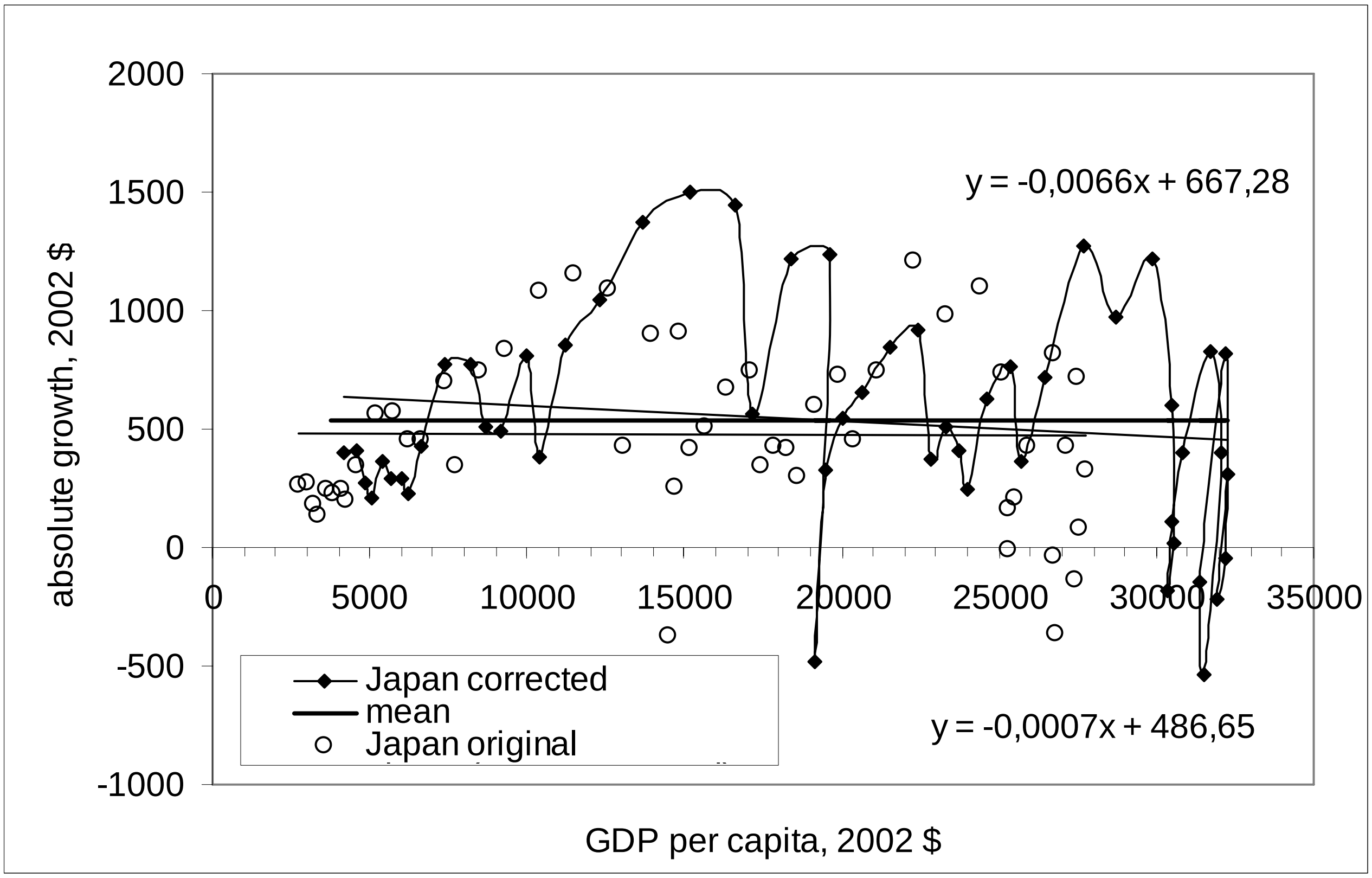

Fig. 13. Same as in Figure 4 for Japan. The mean value is $538.The original linear regression line is parallel to x-axis. The corrected line is characterized by a negative trend. There were two periods of very quick growth between $10000 and $18000 and between $26000 and $31000. Both ended in periods of a low (sometimes - negative) growth rates. Same effect might be expected for Ireland.

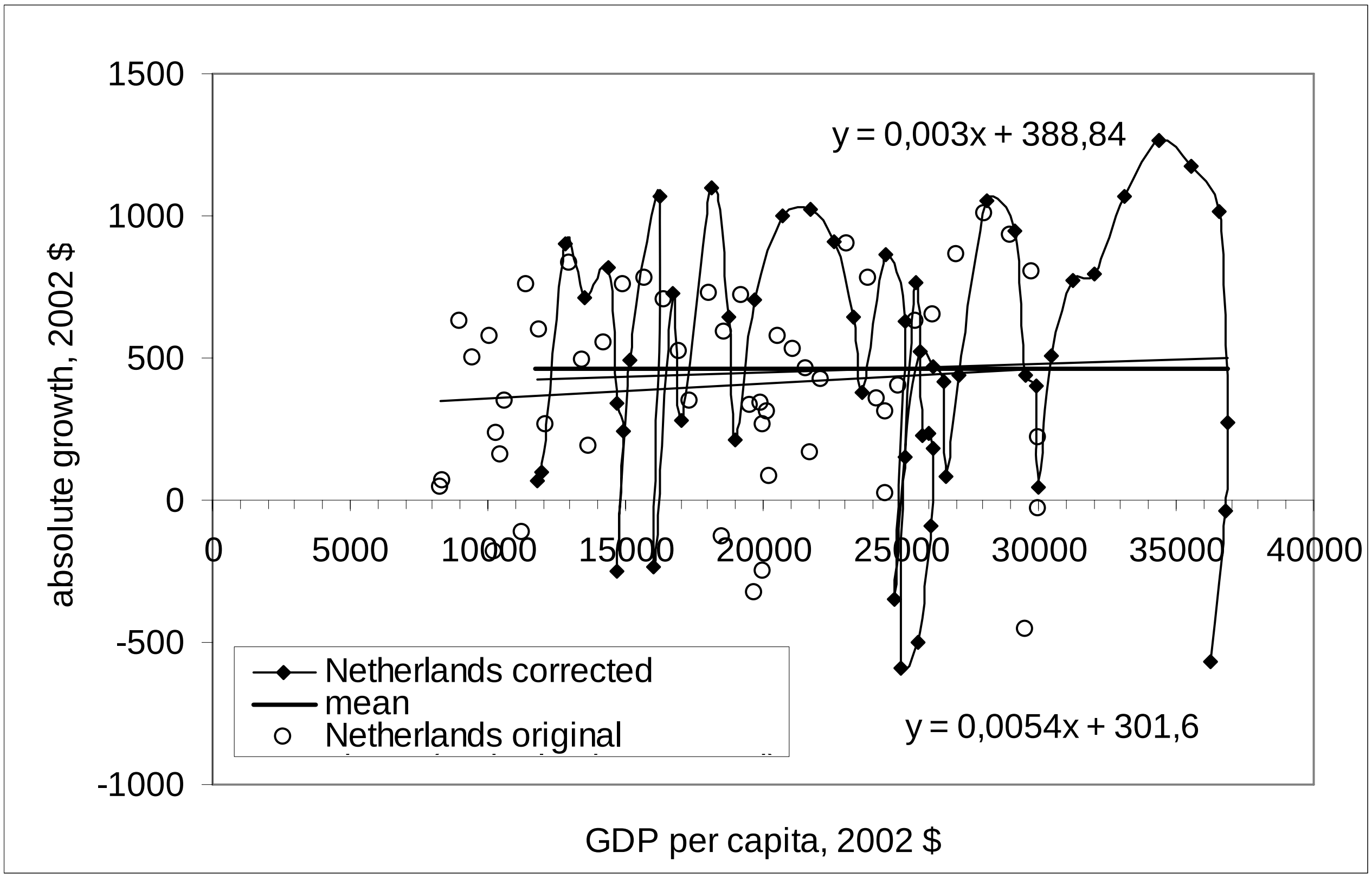

Fig. 14. Same as in Figure 4 for Netherlands. The mean value is $462.

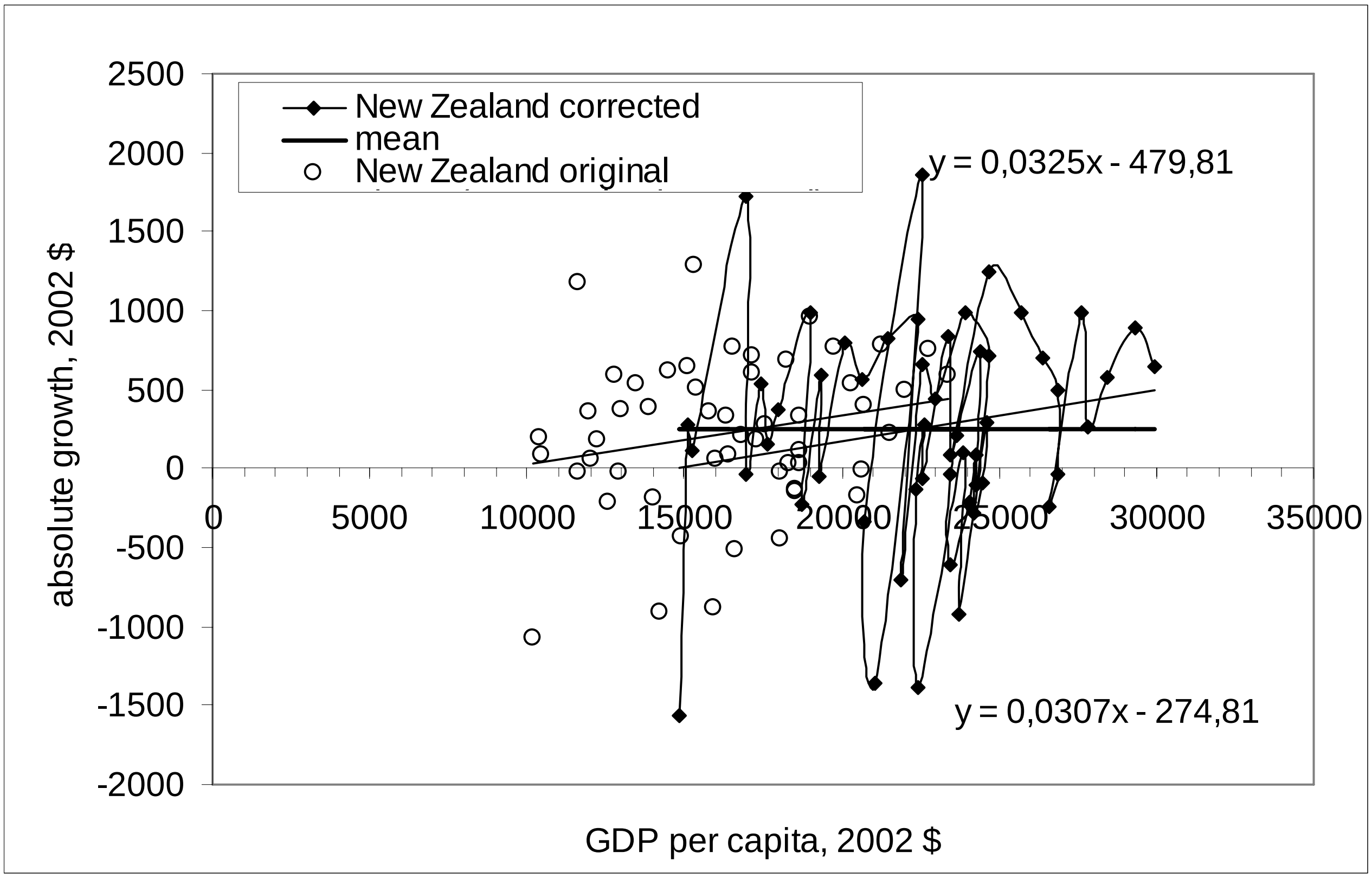

Fig. 15. Same as in Figure 2 for New Zealand. The mean value is very low - $255.

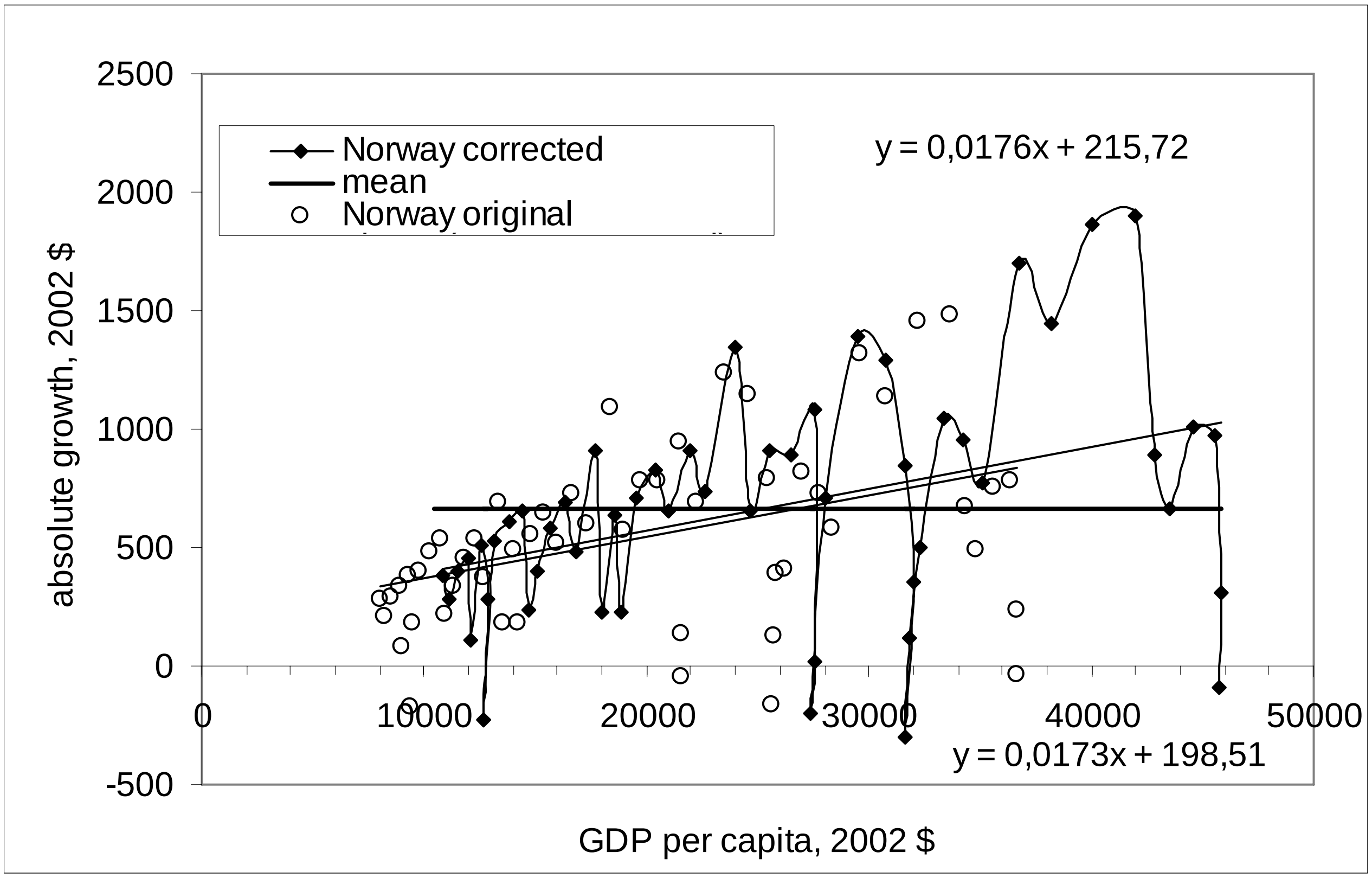

Fig. 16. Same as in Figure 4 for Norway. The mean value is as high as $665.

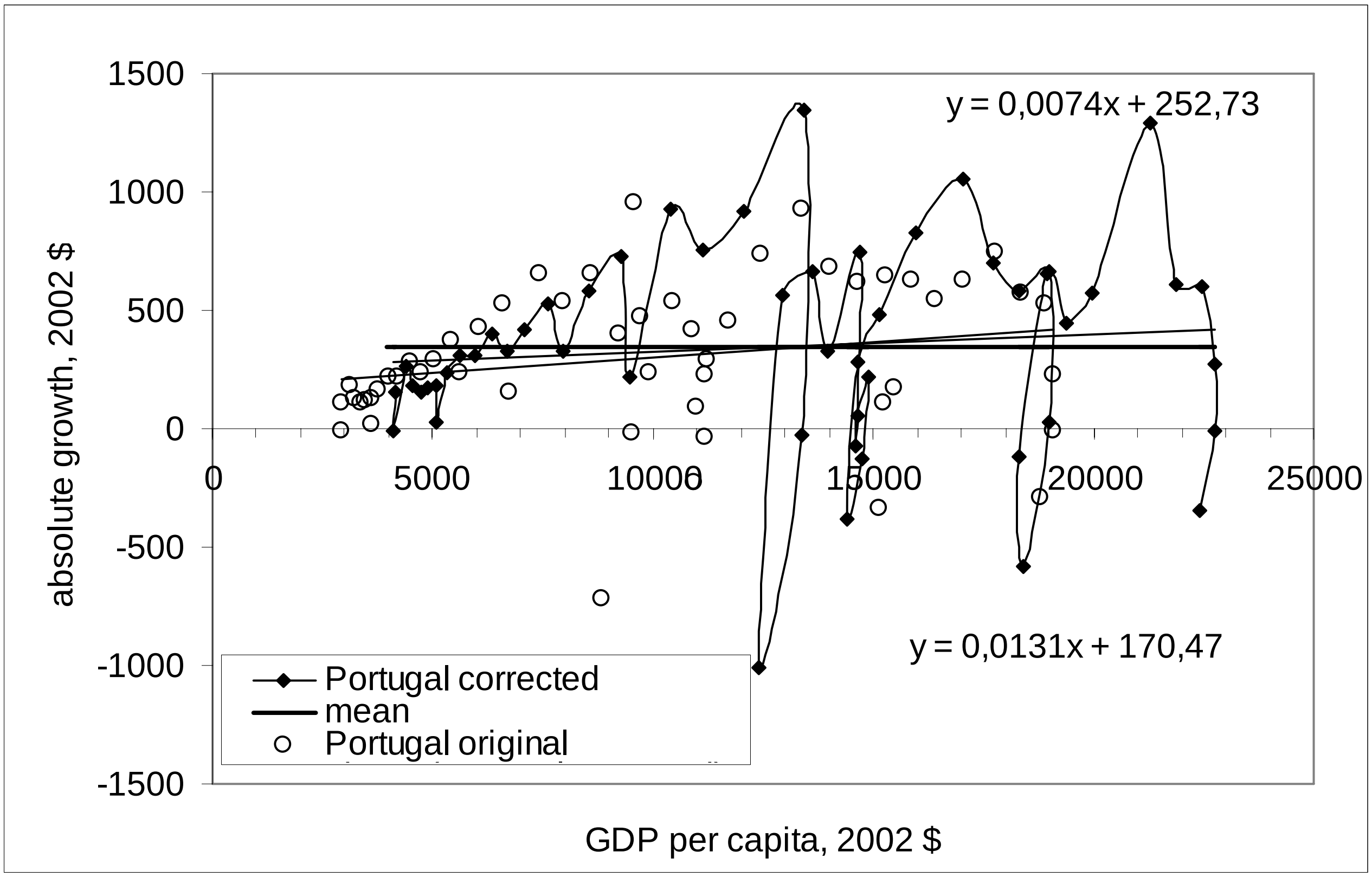

Fig. 17. Same as in Figure 4 for Portugal. The mean value is $348.

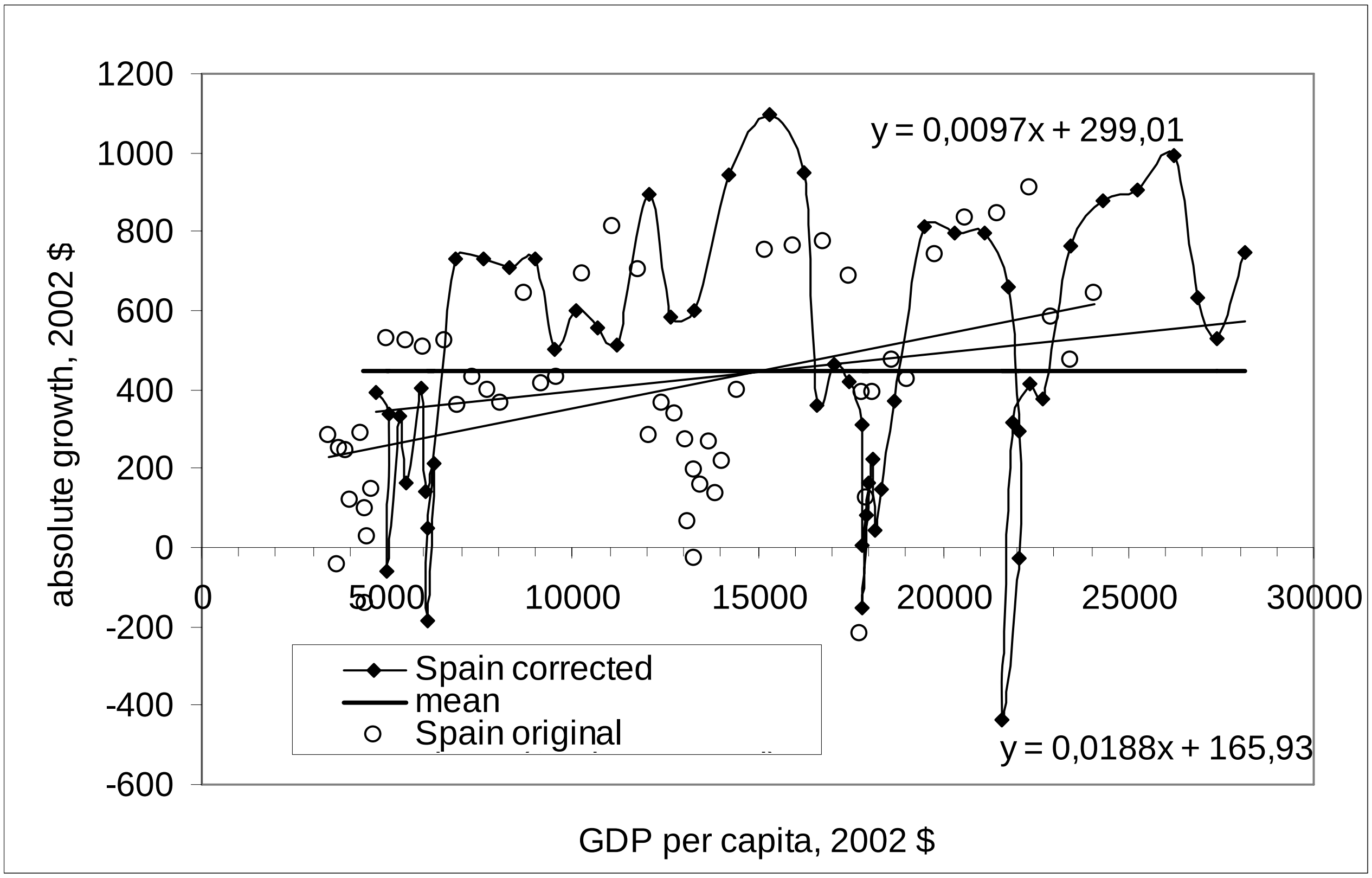

Fig. 18. Same as in Figure 4 for Spain. The mean value is $449. The earlier absolute growth values are relatively low and are compensated during the previous thirty years.

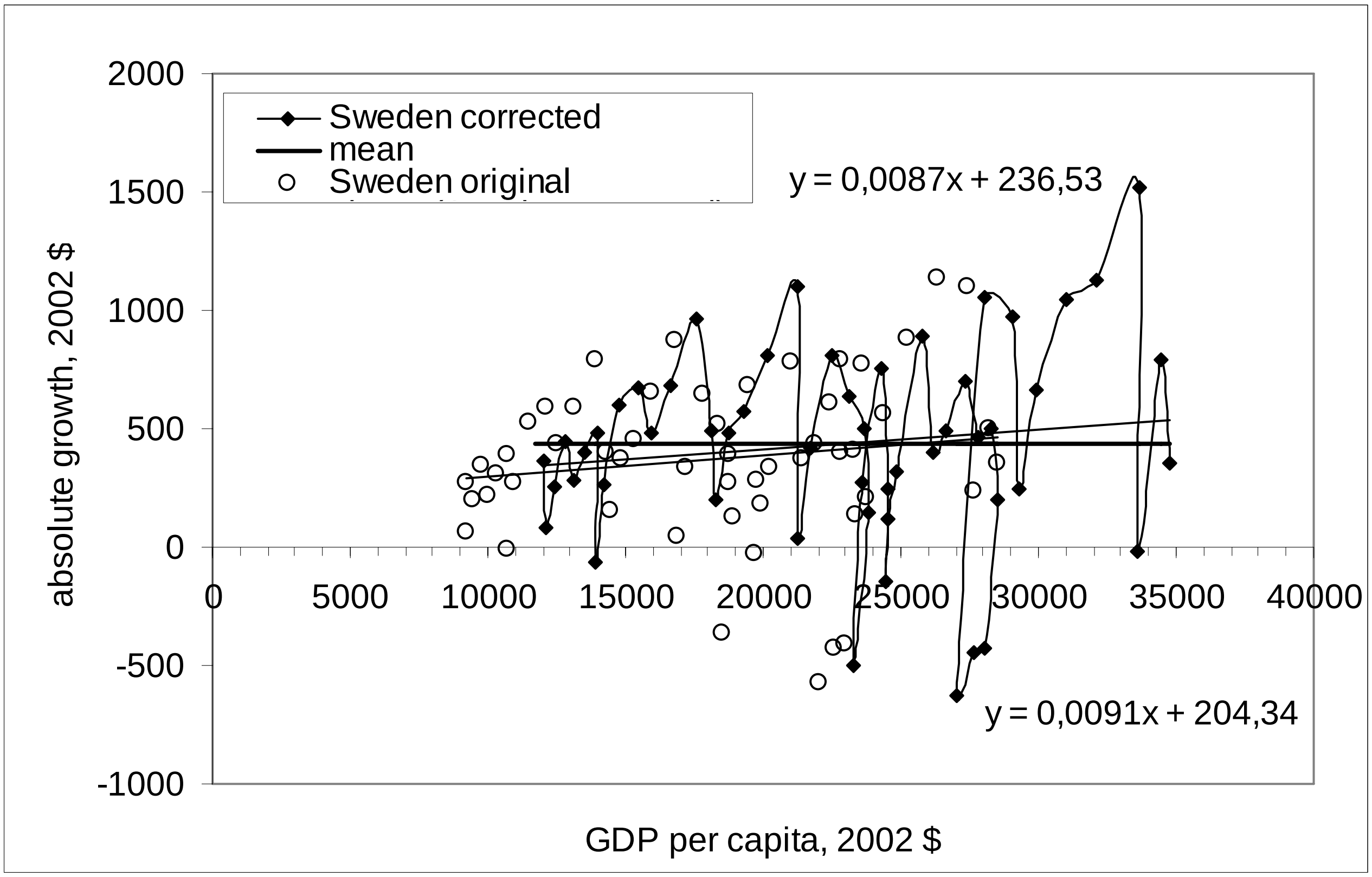

Fig. 19. Same as in Figure 4 for Sweden. The mean value is $436. The linear regression line almost coincides with the mean value straight line.

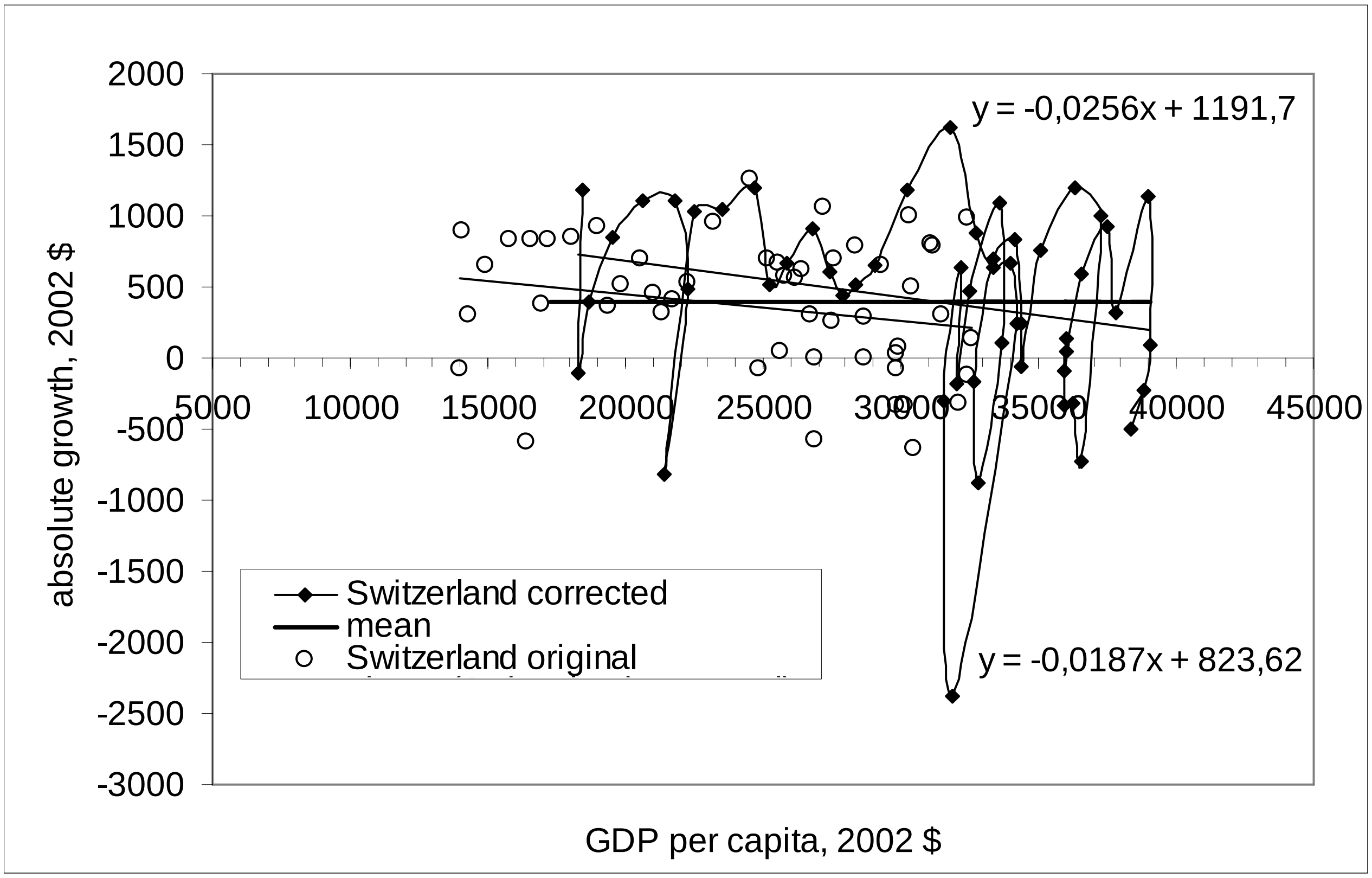

Fig. 20. Same as in Figure 4 for Switzerland. The mean value is $398. The country shows a consistent negative trend in the GDP per capita annual increment.

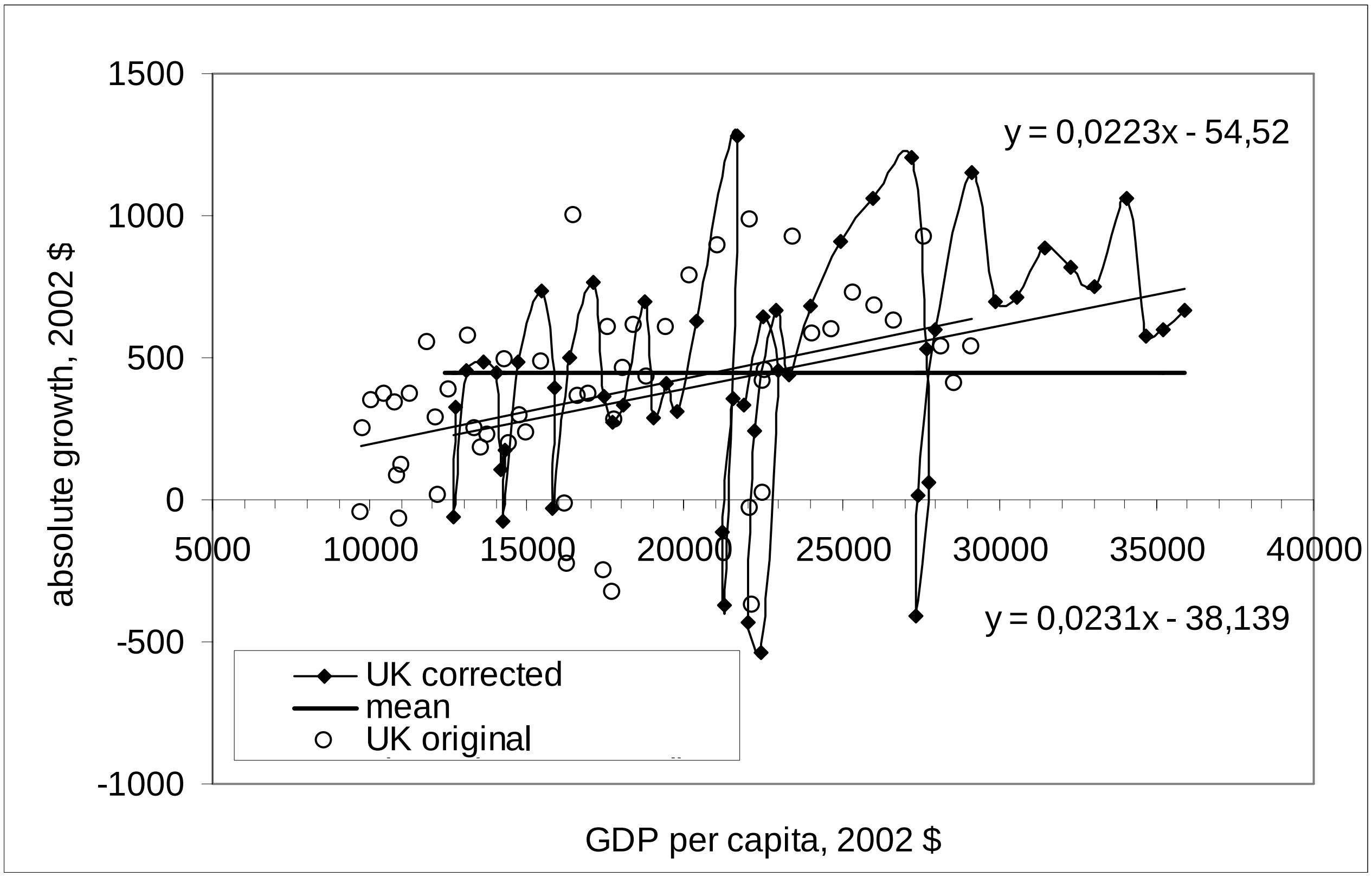

Fig. 21. Same as in Figure 4 for the UK. The mean value is $444. A strong positive trend is observed.

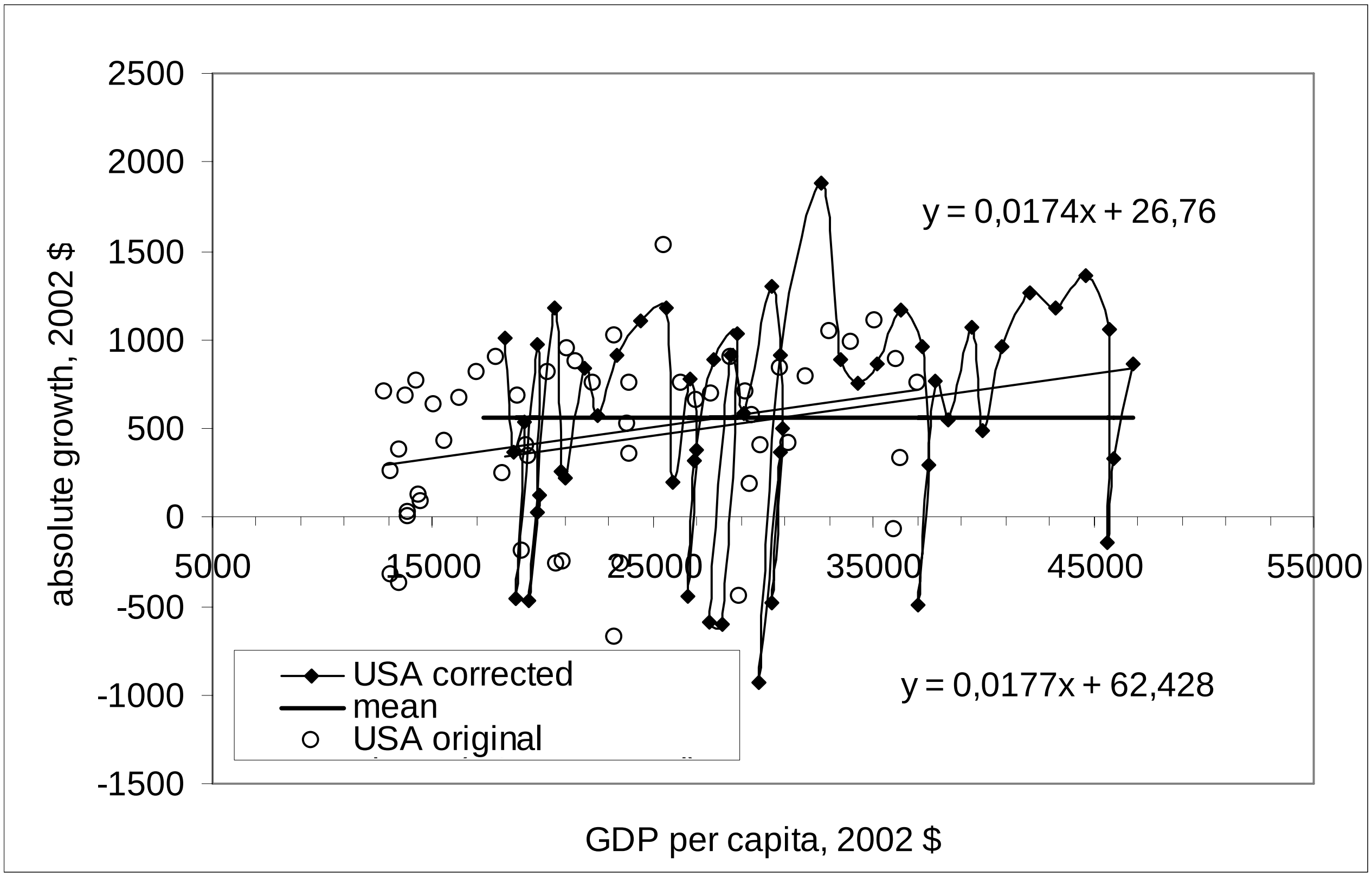

Fig. 22. Same as in Figure 4 for the USA. The mean value is $556.

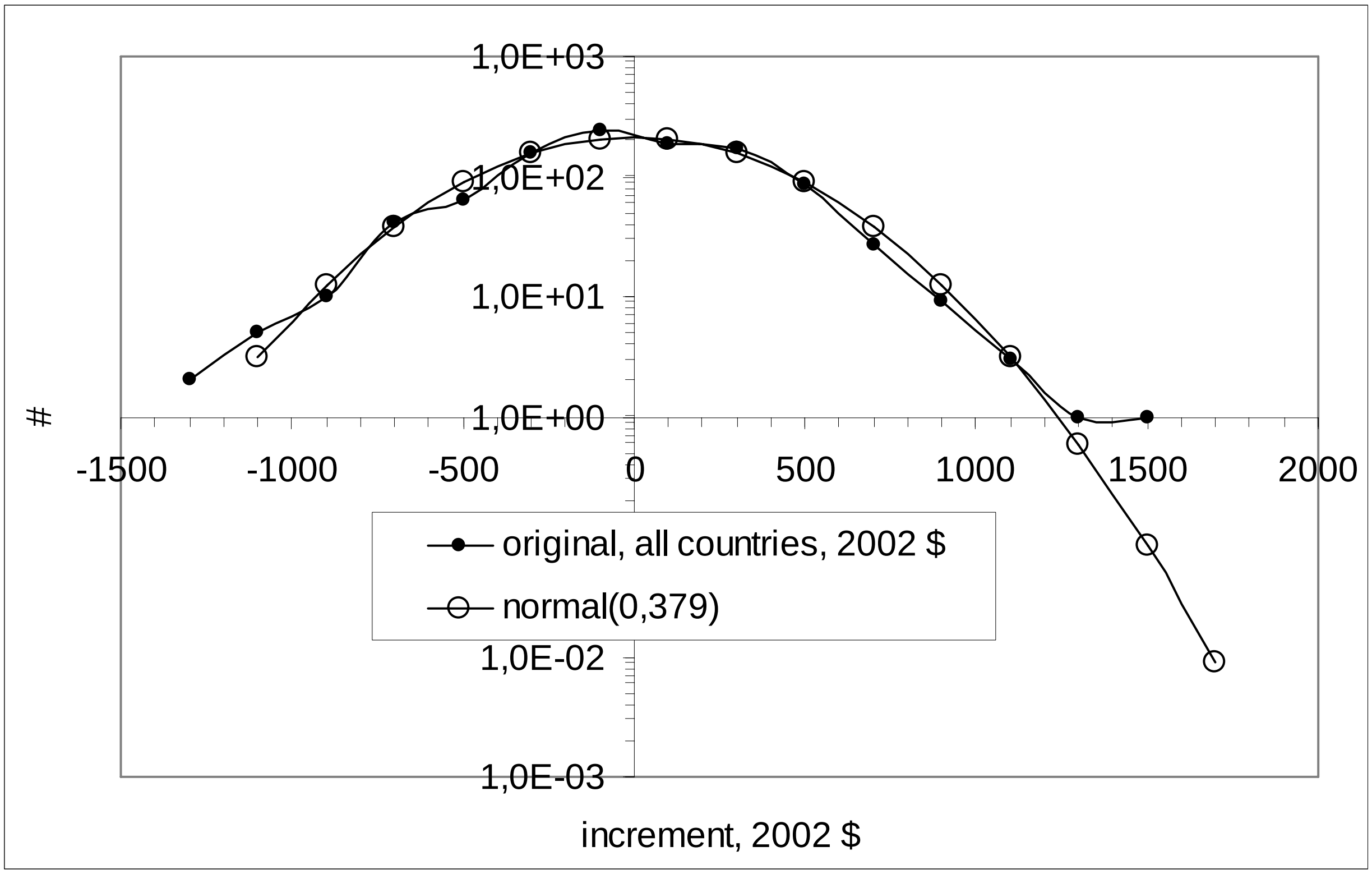

Fig. 23. Frequency distribution of the GDP per capita increment values for the complete (19 countries) original data set. The GDP values are obtained at “EKS” PPPs. The mean values are extracted from the corresponding increment original values resulting in a zero central value of the distribution. Normal distribution with a mean value and standard deviation of the actual distribution is presented by open circles. The normal distribution is very close to the actual one, at least in the central zone.

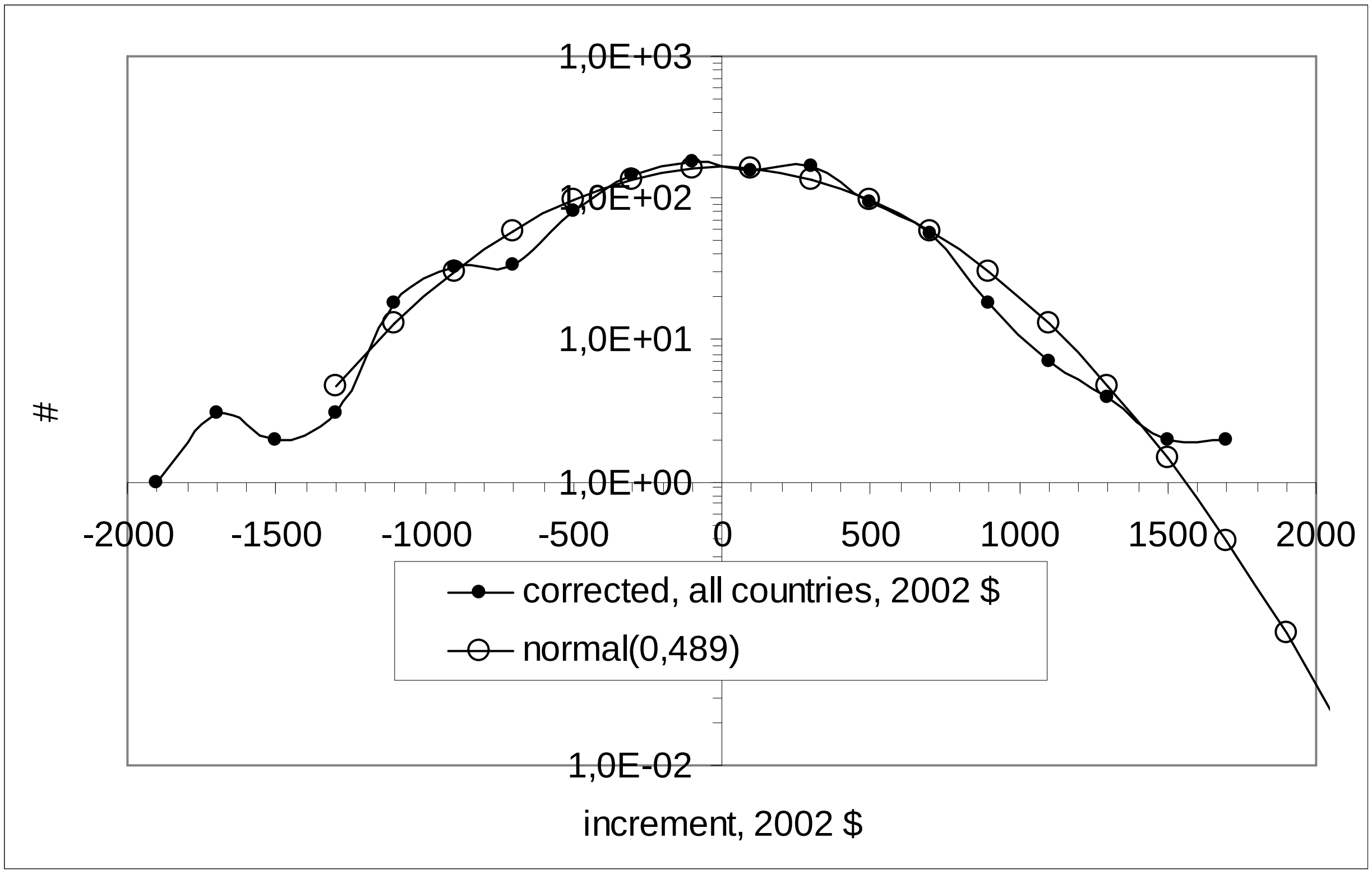

Fig. 24. Same as in Fig. 23 for the population corrected data set. Larger deviations from the normal distribution are observed.

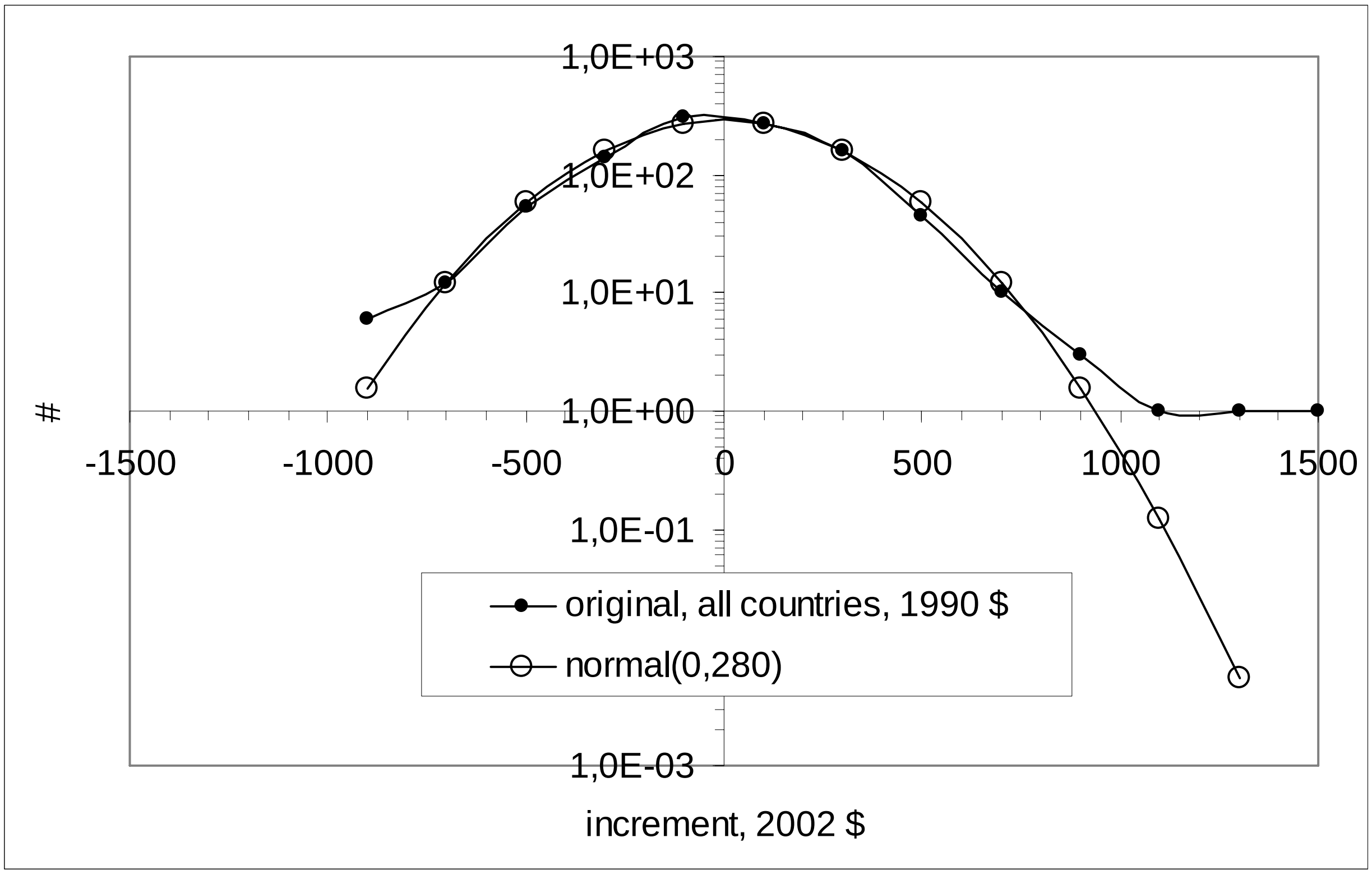

Fig. 25. Same as in Fig. 23, but the GDP values are obtained at “Geary-Khamis” PPPs. The normal distribution describes the actual one better than that obtained for the “EKS” PPPs.

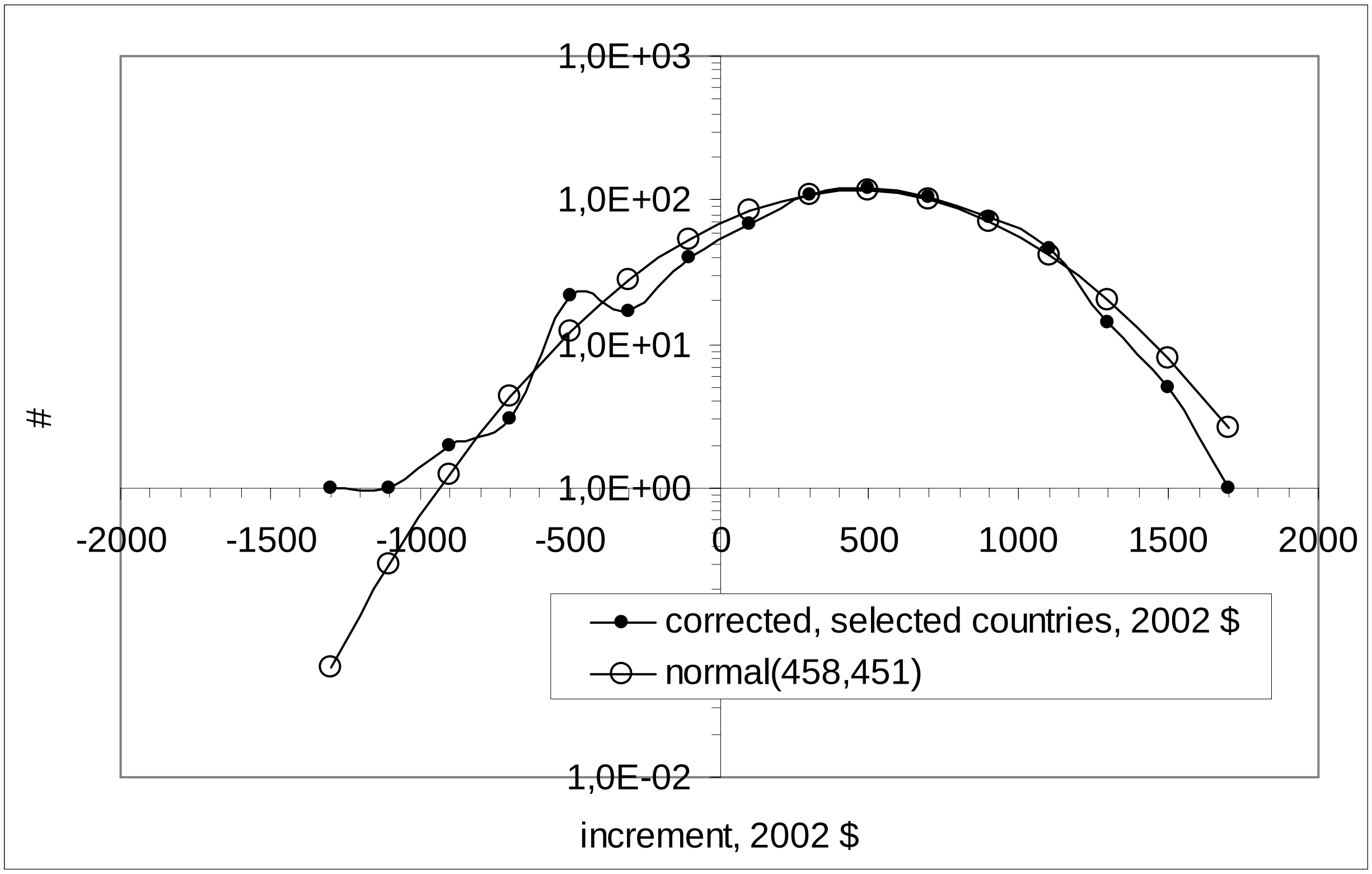

Fig. 26. Same as in Fig. 23, but for the selected countries excluding outliers. The mean value over the countries is $458 and not subtracted from the fluctuations.

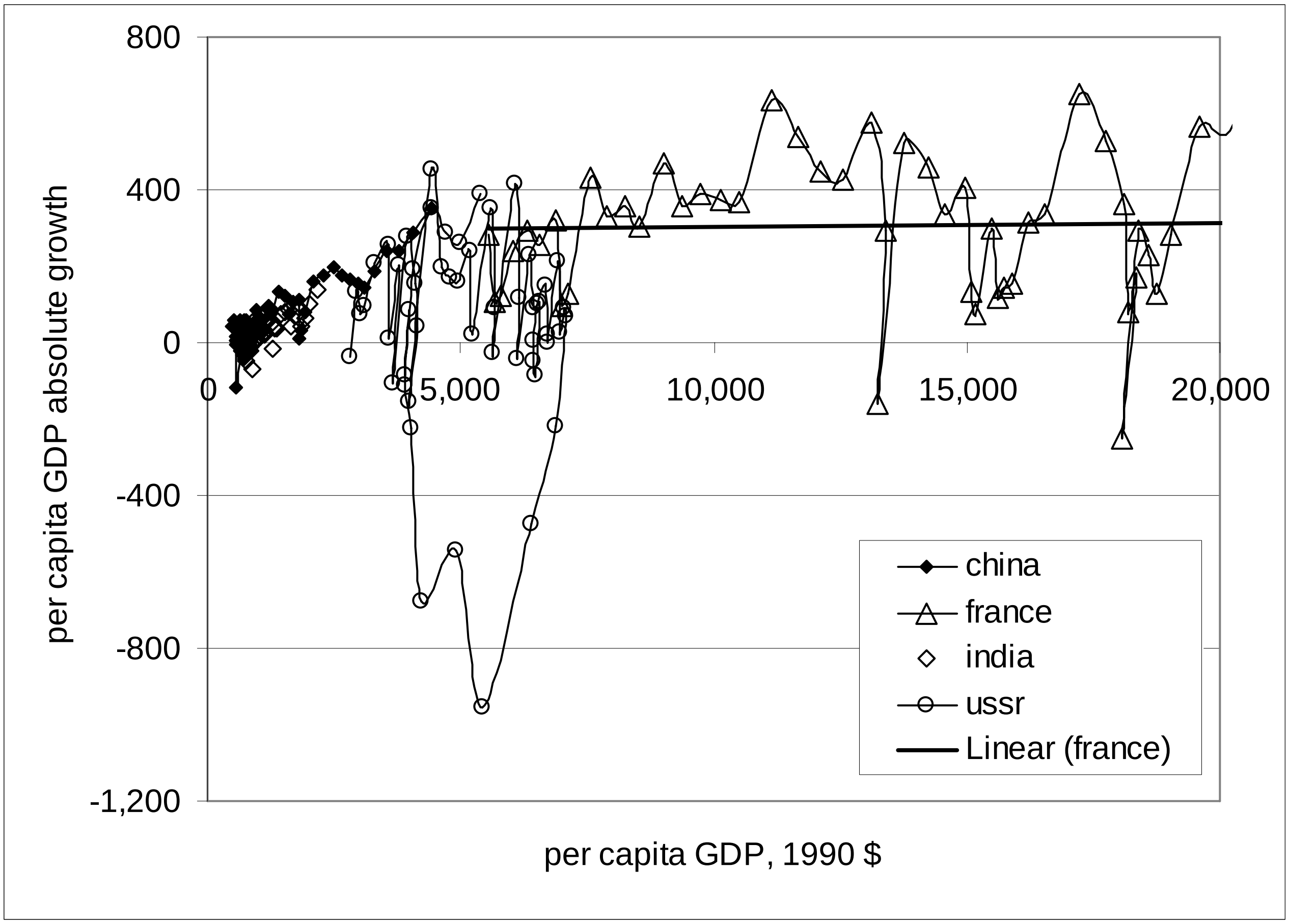

Fig. 27. GDP per capita increment for China, India and the USSR for the period between 1950 and 2003 compared to that for France. GDP is expressed in $ 1990, the only available estimates for the non-OECD countries. India is far below the mean increment for France, but China has just reached the pace of leading developed countries. For the period of existence (between 1950 and 1990 in the study), the USSR was only about a quarter as effective as France.